\documentclass[a4paper]{article}

\usepackage{amsfonts}

\def\be{\begin{equation}}
\def\ee{\end{equation}}
\def\bea{\begin{eqnarray}}
\def\eea{\end{eqnarray}}
\def\({\left(}
\def\){\right)}
\def\<{\left<}
\def\>{\right>}

\def\>{\rangle}
\def\<{\langle}
\def\|{\mid}

\def\tr{{\mbox{tr}}}

\def\be{\begin{equation}}
\def\ee{\end{equation}}
\def\bea{\begin{eqnarray*}}
\def\eea{\end{eqnarray*}}
\def\ben{\begin{eqnarray}}
\def\een{\end{eqnarray}}
\def\({\left(}
\def\){\right)}
\def\<{\left<}
\def\>{\right>}

\def\[{\left[}
\def\]{\right]}

\def\+{\bar}
\def\mb{\mathbb}
\def\tr{{\mbox{tr}}}

\def\D{{\cal{D}}}
\def\L{{\cal{L}}}
\def\H{{\cal{H}}}
\def\t{\tilde}

\def\t{\widetilde}

\def\A{{\cal{A}}}
\def\B{{\cal{B}}}
\def\N{{\cal{N}}}
\def\O{{\cal{O}}}

\def\T{{\cal{T}}}

\def\ee{\underline{e}}

\def\aalpha{\underline{\alpha}}
\def\bbeta{\underline{\beta}}
\def\ggamma{\underline{\gamma}}

\def\G{{\cal{G}}}

\begin{document}
\setlength{\unitlength}{1mm}

\pagestyle{empty}
\vskip-10pt
\vskip-10pt
\hfill 
\begin{center}
\vskip 3truecm
{\Large \bf
Five-dimensional super Yang-Mills theory\\
from ABJM theory}\\
\vskip 2truecm
{\large \bf
Andreas Gustavsson\footnote{a.r.gustavsson@swipnet.se}}\\
\vskip 1truecm
{\it  Center for quantum spacetime (CQUeST), Sogang University, Seoul 121-742, Korea}\\
and\\
\it{School of Physics \& Astronomy, Seoul National University, Seoul 151-747 Korea}
\end{center}
\vskip 2truecm
{\abstract{We derive five-dimensional super Yang-Mills theory from mass-deformed ABJM theory by expanding about $S^2$ for large Chern-Simons level $K$. We obtain the Yang-Mills coupling constant $g_{YM}^2 = 4\pi^2 R/K$. If we consider $S^3/{\mb{Z}_K}$ as a fiber bundle over $S^2$ then $R/K$ is the circumference of the fiber. The value on the coupling constant agrees with what one gets by compactifying M five-brane on that fiber. For this computation we take $R,K\rightarrow \infty$ while keeping $R/K$ at a fixed finite value. We also study mass deformed star-three-product BLG theory at $K=1$ and $R\rightarrow \infty$. In that limit we obtain Lorentz covariant supersymmetry variations and gauge variations of a non-Abelian tensor multiplet.}}

\vfill 
\vskip4pt
\eject
\pagestyle{plain}

\section{Introduction}
In M-theory it appears that three-manifolds take the role of two-manifolds in string theory. Three-algebra underlies the gauge structure of M2 brane theory \cite{G}, \cite{BLG}, \cite{BL} and the Nambu bracket, which is defined on a three-manifold, is one realization. The Nambu bracket does not close on any finite set of functions, except for the case of $SO(4)$ three-algebra, the symmetry group of $S^3$. In all other cases we must include an infinite set of functions to close the algebra. The same is true for the Poisson bracket defined on a two-manifold, but here we have a deformation which is the star-commutator which can be used to obtain finite-dimensional Lie algebras. 

ABJM theory appears as a strong candidate for the theory of M2 branes \cite{ABJM}, \cite{Kim:2010mr}, \cite{Kim:2009wb}, \cite{Bashkirov:2010kz}, \cite{Kapustin:2010xq}, \cite{Drukker:2010nc}. But if we mass-deform ABJM theory, we can not see $S^3$ as a classical solution to the field equations, but only the $S^2$ base-manifold \cite{Nastase:2010uy}. ABJM theory might describe all aspects of M2 branes correctly as a quantum theory. However it is hard to use ABJM theory to deconstruct the M5 brane wrapped on $S^3$ by fluctuation analysis, unless $S^3$ is a classical solution to the M2 brane theory. So we wish to seek a theory where $S^3$ is a classical solution. If $S^3$ is a solution to quantum ABJM theory, then perhaps one may think on this new theory as a quantum effective theory of ABJM theory. 

In this paper we will derive the D4 brane Lagrangian from ABJM theory. More specifically we will obtain D4 wrapped on $S^2$ base-manifold of $S^3/{\mb{Z}_K}$. We will obtain the same answer including the right value on the Yang-Mills coupling constant \cite{Tong:2005un}, 
\ben
g_{YM}^2 = 4\pi^2 \frac{R}{K}\label{coupl}
\een
as if we had dimensionally reduced M5 brane on the fiber of $S^3/{\mb{Z}_K}$. For this it is necessary to identify $K$ as the Chern-Simons level in ABJM theory. The emergence of the M5 brane coupling constant from ABJM theory is a remarkable result. It gives us hope that M5 brane physics can be extracted from ABJM theory. 

In \cite{Nastase:2010uy} a single D4 is obtained from ABJM theory. As this is a free theory we do not have a coupling constant. If nevertheless we would identify the overall factor multiplying the Lagrangian as a coupling constant, we would realize that a rescaling of the fields would change that overall constant. We can only determine the coupling constant if we require the gauge field to be normalized to satisfy the Dirac charge quantization condition. But this does not seem to be taken care of in \cite{Nastase:2010uy}, and we will not attempt this here either. Instead we will deconstruct non-Abelian D4 theory where we can safely identify $g_{YM}$.

In section \ref{Three-algebra} we study the three-algebra of functions on $S^3$ modulo an error of order $1/N$. In section \ref{The D4 brane} we write down the D4 brane action and Abelian M5 and relate their coupling constants. In section \ref{ABJM theory} we write down ABJM theory with its mass deformation. In section \ref{Maximally supersymmetric vacuum} we obtain the Bogomolnyi equation for supersymmetric vacua. In section \ref{D4 from ABJM} we deconstruct D4 from ABJM and obtain (\ref{coupl}). In section \ref{Lagrangian for a selfdual three-form} we discuss a Lagrangian for a selfdual three-form. In section \ref{Multiple M5 in decompactification limit} we partly derive from ABJM, and partly guess, the gauge variations and supersymmetry variations for what we believe is multiple M5 theory. 
 
After the main part of this paper had been completed, a paper \cite{Terashima:2010ji} appeared which appears to have conceptual overlaps with our work. Other recent works are \cite{Lambert:2010iw}, \cite{Douglas:2010iu}.

\section{Three-algebra}\label{Three-algebra}
We can describe ABJM theory as a usual gauge theory. We have a Lie algebra associated with the gauge group, and the fields transform in certain representations of the gauge group. Three-algebra is not needed to describe ABJM theory. On the other hand it is possible to describe ABJM theory in the language of hermitian three-algebra \cite{BL}, \cite{Gustavsson:2010yr}. One realization of three-algebra is the Nambu three-bracket. This realization is hard to come by if one uses the Lie algebra language. To each three-algebra there is an associated ABJM theory. Only certain Lie algebras give ABJM theories \cite{Schnabl:2008wj}, \cite{Gustavsson:2010yr}.

\subsection{Three-product, Nambu-bracket and dimensional reduction}
We will assume that we have some three-associative three-product,
\bea
(T^a T_c T^b) T_d T^e = T^a (T_c T^b T_d) T^e = T^a T_c (T^b T_d T^e).
\eea
where $T^a$ denote elements in a three-algebra. We assume the existence of a conjugation, that is an operation '*' that squares to one. We denote by $T_a = (T^a)^*$ the conjugate elements, and we thus have $(T_a)^* = (T^a)^{**} = T^a$. We define the three-bracket of three three-algebra elements $T^a$, $T^b$ and $T^c$ as
\ben
[T^a,T^b;T^c] &=& T^a T_c T^b - T^b T_c T^a\label{abstract}
\een 
which is a map from three three-algebra elements into a new three-algebra element. As seen from the definition, the three-bracket is complex anti-linear in its third entry. At this stage we do not make any assumption as to whether the conjugate elements $T_a$ belong to the  three-algebra or not. They may or may not be related to elements in the three-algebra. We do not include the conjugate elements as additional possibly independent elements in the set of three-algebra elements. That is the reason we choose to denote the three-bracket as $[T^a,T^b;T^c]$. This is the same notation as used in \cite{Cherkis:2008ha}, but other notations for the same bracket has also appeared in the literature.\footnote{The most commonly used notation appears to be $[T^a,T^b;T_c]$ for this three-bracket. This notation makes indices on the left and right-hand sides of equation (\ref{abstract}) be positioned at the same level. This notation is confusing in this context though. First it is incompatible with assuming the bracket is complex anti-linear in the third entry. Second the notation makes the notion of three-algebra elements obscure as the $T_a$ are not elements of the algebra.} It can be shown that (\ref{abstract}) satisfies the hermitian fundamental identity
\ben
[[T^a,T^b;T^c],T^e;T^f] &=& [[T^a,T^e;T^f],T^b;T^c] + [T^a,[T^b,T^e;T^f];T^c]\cr
&& - [T^a,T^b;[T^c,T^f;T^e]].\label{fi}
\een
if conjugation acts on the three-product as
\bea
(T^a T_c T^b)^* &=& T_b T^c T_a.
\eea
Then conjugation acts on the tree-bracket with a minus sign as
\ben
[T^a,T^b;T^c]^* &=& -[T_a,T_b;T_c]\label{conj} 
\een
We also find that the following Leibniz rule is obeyed,
\ben
[T^a T_c T^b,T^e;T^f] = [T^a,T^e;T^f]T_c T^b + T^a[T_c,T_f;T_e]T^b + T^a T_c [T^b,T^e;T^f].\label{Leibniz}
\een
Under certain circumstances the three-product realization of the three-bracket can also be expressed as
\ben
[T^a,T^b;T^c] &=& T_c[T^a,T^b] + [T^a,T_c]T^b - [T^b,T_c]T^a\label{surprise}
\een
where a quantity like $[T^a,T^b]$ by itself is generically not defined as this involves only two elements, but quantities like $T_c[T^a,T^b] = T_cT^a T^b - T_c T^b T^a$ may be well-defined in terms of the three-product. One realization of the three-product is by means of matrices. In this case $T_a = (T^a)^{\dag}$ is given by the hermitian conjugate matrix, and $[T^a,T^b]$ is a well-defined matrix commutator as long as we restrict ourselves to $N\times N$ square matrices. If we assume generic $N\times M$ matrices $T^a$ and $M\times N$ hermitian conjugate matrices $T_a$, there is generically no notion of a product $T^a T^b$ but we may still consider products of matrices on the alternating form $T^a T_b T^c T_d T^e ...$. 

One solution to the fundamental identity (\ref{fi}) is provided by the Nambu bracket,
\bea
[\T^a,\T^b;\T^c] &=& \hbar \{\T^a,\T^b,\T_c\},\cr
\{\T^a,\T^b,\T_c\} &=& \sqrt{g}\epsilon^{\aalpha\bbeta\ggamma} \partial_{\aalpha}\T^a \partial_{\bbeta}\T^b \partial_{\ggamma}\T_c\label{Namburealization}
\eea
where 
\bea
\sigma^{\aalpha} &=& (\sigma^m,\sigma^3)
\eea
are three coordinates on $\mb{R}^3$ with metric $ds^2 = g_{\aalpha\bbeta} d\sigma^{\aalpha} d\sigma^{\bbeta}$ and the determinant of the metric is denoted as $g$. We use the convention where $\epsilon_{123} = 1$ and indices are rised by the inverse metric, $g^{\aalpha\bbeta}$. 

If we restrict ourselves to functions on the form 
\ben
\T^a &=& e^{i\sigma^3} \t \T^a(\sigma^{m})\label{red}
\een
the Nambu three-bracket reduces as
\ben
[\T^a,\T^b;\T^c]  &=& -i\hbar \sqrt{\frac{G}{g}} \(\T_{c}\{\T^{a},\T^{b}\} + \{\T^{a},\T_{c}\}\T^{b} - \{\T^{b},\T_{c}\}\T^{a}\).\label{reduced}
\een
where $\{\T^a,\T^b\} = \sqrt{G}\epsilon^{mn}\partial_m \T^a \partial_n \T^b$ and the determinant is denoted as $G$. We will refer to this as 'dimensional reduction' of the Nambu bracket to Poisson brackets. We note that (\ref{reduced}) is on the same form as (\ref{surprise}). This is surprising since the Nambu bracket was defined with no mention of a three-product. The structure of the Nambu bracket appears to be quite different from the three-product structure of the three-bracket defined in (\ref{abstract}).

Matrix commutator is mapped isomorphically into star-commutator. To lowest order in the purely imaginary two-dimensional  non-commutativity parameter $\epsilon$, we have
\ben
[T^a,T^b] &\cong & \epsilon \{\T^a,\T^b\}.\label{iso}
\een
where we have the matrix commutator in the left-hand side. We will let
\ben
\epsilon &=& -i\hbar\sqrt{\frac{G}{g}}\label{epsilon}
\een
where $\hbar$ is the real three-dimensional non-commutativity parameter. The non-commutativity parameter $\epsilon$ must be purely imaginary, because only then we find the same minus signs on both sides of (\ref{iso}) under conjugation. On the LHS we get $[T^a,T^b]^{\dag} = -[T_a,T_b]$ because hermitian conjugation acts on a product of matrices as $(T^a T^b)^{\dag} = T_b T_a$, on the RHS we get $(i\{\T^a,\T^b\})^* = -i \{\T_a,\T_b\}$ because of the factor of $i$. For the Poisson bracket conjugation acts just as expected, $\{\T^a,\T^b\}^* = \{\T_a,\T_b\}$. Likewise for the Nambu bracket we have $\{\T^a,\T^b,\T_c\}^* = \{\T_a,\T_b,\T^c\}$. Under dimensional reduction it is mapped into (\ref{reduced}) divided by $\hbar$. If we take the complex conjugate it appears like we would have $\{\T^a,\T^b,\T_c\}^* = -\{\T_a,\T_b,\T^c\}$, but that is not true. We are then forgetting that the form of the dimensionally reduced Poisson brackets in (\ref{reduced}) depend on the phase factor $e^{i\sigma^3}$ in the $\T^a$'s. Under complex conjugation of the Nambu bracket, these phase factors are conjugated into $e^{-i\sigma^3}$, whereby we get dimensionally reduced Poisson brackets with opposite sign. Taking this additional sign change into account, everything is consistent with assuming that $\{\T^a,\T^b,\T_c\}^* = \{\T_a,\T_b,\T^c\}$.

\subsection{Star-three-product}
There does not seem to exist a deformation of the Nambu bracket on the form
\bea
[\T^a,\T^b;\T^c] &=& \hbar \{\T^a,\T^b,\T_c\} + \O(\hbar^2)
\eea
which still satisfies the fundamental identity \cite{Chen:2010ny}. Only the linear term, which is the Nambu bracket, satisfies the fundamental identity. To go beyond this, we include additional terms at linear order, on the form
\bea
[\T^a,\T^b;\T^c] &=& \hbar \(\{\T^a,\T^b,\T_c\} - \T_c\{\T^a,\T^b,\bullet\} + 2 \T^{[a}\{\T^{b]},\T_c,\bullet\}\) + \O(\hbar^2)
\eea
for certain coefficients $a$ and $b$. An associative star-three-product that gives a star-three-bracket on form above, was found in \cite{Gustavsson:2010nc} and is given by
\bea
\T^a * \T_c * \T^b(\sigma) &=& \lim_{\sigma \rightarrow \sigma' \rightarrow \sigma''}\exp \Big\{\frac{\hbar}{2}\sqrt{g}\epsilon^{\aalpha\bbeta\ggamma}\cr
&&\(\partial_{\aalpha}\partial_{\bbeta}''\partial_{\ggamma}'+\partial_{\aalpha}''\partial_{\bbeta}'\partial_{\ggamma}^{out}-\partial_{\aalpha}\partial_{\bbeta}'\partial_{\ggamma}^{out}-\partial_{\aalpha}\partial_{\bbeta}''\partial_{\ggamma}^{out}\) \Big\}\cr
&&\T^a(\sigma)\T_c(\sigma')\T^b(\sigma'')
\eea
The outer derivatives $\partial_{\aalpha}^{out}$ act on the next star-three-product by which this star-three-product may be three-multiplied, for example as $(\T^a * \T_c * \T^b) * \T_e * \T^d$ but still we have some outer derivatives surviving, no matter how many nested star-three-products we consider. Due to these outer derivatives we do not get a closed three-algebra on any finite or infinite set of harmonic functions or any other basis of functions. We must extend the set of generators from functions $\T^a$ to function-differential-operators on the form $\T^a_A = \T^a \otimes \D_A$. On a three-torus with euclidean metric, we may take generators as $\T^m_M = \T^m \otimes \D_M$ with 
\bea
\T^m &=& e^{im_{\aalpha}\sigma^{\aalpha}},\cr
\D_M &=& e^{-\frac{\hbar}{2}M^{\aalpha}\partial_{\aalpha}^{out}}
\eea
where $m_{\aalpha}$ and $M^{\aalpha}$ are integers. We know how complex conjugation acts on $\T^m$. We shall of course have $(\T^m)^* = e^{-im\sigma}$. Slightly more tricky is the complex conjugation of $\D_M$. It should not be confused with hermitian conjugation with respect to the natural inner product on function space, where we have $\partial^{\dag} = - \partial$. Here we rather shall take $\partial^* = \partial$. As motivation for this, let us consider the complex conjugation of $\partial e^{im\sigma}$. This is given by $\partial e^{-im\sigma}$ and not by $(-\partial)e^{-im\sigma}$. To summarize then, we shall have
\bea
(\T^m_M)^* & \equiv & \T_m^M \cr
&=& e^{-im_{\aalpha}\sigma^{\aalpha}} \otimes e^{-\frac{\hbar}{2} M^{\aalpha}\partial_{\aalpha}^{out} }.
\eea
With these assignments we obtain
\bea
T^m_M * T_p^P * T^n_N &=& e^{\frac{i\hbar}{2}\varphi(m,M,n,N,p,P)} T^{m+n-p}_{M+N+P-X}
\eea
where
\bea
\varphi(m,M,n,N,p,P) &=& \{m,n,p\} - M(n-p) - N(m-p) - P(m+n),\cr
X^{\aalpha} &=& -\sqrt{g} \epsilon^{\aalpha\bbeta\ggamma} \(n_{\bbeta}p_{\ggamma} - m_{\bbeta}p_{\ggamma} + m_{\bbeta}n_{\ggamma}\)
\eea
Here 
\bea
\{m,n,p\} &=& -\frac{i\hbar}{2}\epsilon^{\aalpha\bbeta\ggamma}m_{\aalpha}n_{\bbeta}p_{\ggamma}.
\eea
If we choose $\hbar = \frac{4\pi}{N}$, then the structure constants will be invariant under $m_{\aalpha} \rightarrow m_{\aalpha} + N$. If we act with these outer derivatives only on functions $\T^m$ for $m=0,...,N-1$, then we see that $\D_{M^{\aalpha}} = \D_{M^{\aalpha}+N}$. In that case, we have a finite-dimensional three-algebra generated by $\T^m_M$ where both $m_{\aalpha}$ and $M^{\aalpha}$ run over the finite set $\{0,...,N-1\}$. Being finite-dimensional, it ought to have a matrix realization in the classification of \cite{Schnabl:2008wj}.

To get an idea how one may obtain an invariant trace form, it seems natural to try to up-lift the situation in two dimensions. In two dimensions, an invariant trace form can be constructed using the Leibniz rule
\bea
[T^a T^b,T^c] &=& [T^a,T^c]T^b + T^a [T^b,T^c] 
\eea
of a (star- or matrix-) commutator. Then we can take the trace
\bea
\tr([T^a T^b,T^c]) &=& \tr([T^a, T^c]T^b) + \tr(T^a[T^b,T^c]) 
\eea
and by the cyclicity of the trace the LHS vanishes, and we obtain the invariance condition for the trace form $\<T^a,T^b\> = \tr(T^a T^b) = \int T^a * T^b$. In three dimensions the analogous thing to consider would be a trace form of three elements,
\bea
\<\T^m_M,\T^n_N,\T^p_P\> &=& \int \frac{d^3 \sigma}{(2\pi)^3} \T^m_M * \T_n^N * \T^p_P.
\eea
The invariance condition 
\bea
\<[\T^m_M,\T^r_R;\T^s_S],\T^n_N,\T^p_P\> + \<\T^m_M,[\T^n_N,\T^s_S;\T^r_R],\T^p_P\> + \<\T^m_M,\T^n_N,[\T^p_P,\T^r_R;\T^s_S]\> &=& 0
\eea
follows from the Leibniz rule together with the fact that 
\bea
\<[\T^m_M,\T^n_N;\T^p_P]\> &=& 0
\eea
The latter property follows from that the integrand, which is a three-bracket, is a total derivative.\footnote{We note that for three matrices we do not seem to have a corresponding property,
\bea
\<[T^a,T^b;T^c]\> = \tr (T^a T_c T^b - T^b T_c T^a) \neq 0.
\eea
It now seems impossible to map our star-three-product isomorphically into matrices.}

But it is not desired to work with a trace form with three entries if we want to use it in a Lagrangian. This is a first signal that something bad could be going on. We could try and define an invariant inner product of two elements as
\bea
\<\T^a_A,\T^b_B\> &=& \<\T^a_A,\T^b_B,1\>.
\eea
But this inner product does not look very natural. Anyway it is invariant, so let us elaborate on this inner product a bit more. Explicitly on the three-torus we obtain
\bea
\<\T^m_M,\T^n_N\> &=& e^{\frac{i\hbar}{2}(M^{\alpha}n_{\alpha}-N^{\alpha}m_{\alpha})} \delta_{m-n,0}.
\eea

Let us now compute the inner product for an interaction term. We start by computing 
\bea
\<\T^m_M * \T_p^P * \T^n_N,\T^q_Q\> &=& e^{\frac{i\hbar}{2} \(Mm+Nn-Pp-Qq\)} \delta_{m+n-p-q,0} 
\eea
and through a cancellation, there is no piece surviving in the exponent which is antisymmetric under exchange of the pair $m,M$ with $n,N$, such as $\{m,n,p\}$. These terms all cancel. This means that 
\bea
\<[\T^m_M,\T^n_N;\T^p_P],\T^q_Q\> &=& 0
\eea
and consequently all interaction terms in ABJM theory that one would build based on star-three-product using this (trivially) invariant inner product, will vanish and we end up with just a free theory. 

As we will see later, all is not lost however. Star-three-product formalism is still a very useful tool to describe ABJM theory. But we will need to use a different inner product, as will be presented in Eq (\ref{inner}). This inner product is trace invariant if we restrict ourselves to generators on the form of Eq (\ref{red}).

\subsection{Truncation}
If we assume that $\T^a = e^{i\sigma^3} \t \T^a$ (dimensional reduction), then we get \cite{Gustavsson:2010nc}
\bea
(\T^a * \T_c * \T^b) \cdot 1 &=& \T^a *_2 \T_c *_2 \T^b,\cr
(\T^a * \T_c * \T^b * \T_e * \T^d) \cdot 1 &=& \T^a *_2 \T_c *_2 \T^b *_2 \T_e *_2 \T^d.
\eea
Here we indicate by acting on the constant function $1$ that all differential operators are dropped from the star-three-product. On the right-hand side we have the two-dimensional star-products. We will refer to this procedure as 'dimensional reduction' plus 'truncation'. By dimensional reduction we refer to that we reduce the star-three-product to two consequtive star-products in two dimensions. By truncation we refer to the action on $1$. This is a consistent truncation only after dimensional reduction, in the sense that we end up with an associative star-product. If we do the truncation on the star-three-product then we loose associativity. But associativity is recovered after dimensional reduction. We have the following commuting diagram 
\bea
\begin{array}{ccc}
{\mbox{star-three-product}} & \rightarrow & {\mbox{truncated star-three-product}}\\
\downarrow &     & \downarrow\\
{\mbox{reduced star-three-product}} & \rightarrow & {\mbox{star-product}}
\end{array}
\eea
where horizontal arrows mean truncation and vertical arrows mean dimensional reduction. Only the upper left and lower right corners correspond to associative products. The other two corners do not. The reduced star-three-product refers to what we get by dimensionally reducing the star-three-product without acting on $1$ to kill the derivatives. This dimensionally reduced product is not associative in the two-dimensional sense. These two non-associative corners can be thought of as intermediate steps between the associative three- and two-dimensional star-products respectively.

The star-three-bracket is not totally antisymmetric due to the outer derivatives. If we could kill these, by for instance acting on the constant function which is equal to $1$, then we obtain a totally antisymmetric truncated three-bracket, which however in general will not satisfy the hermitian (or real) fundamental identity. 

If we assume that all three-algebra elements carry the same phase factor along the fiber according to Eq (\ref{red}), then the remarkable thing happens that the truncated three-bracket which is totally antisymmetric three-bracket, does satisfy the hermitian fundamental identity. The quick way to see this is by expanding out the truncated three-bracket in terms of star-(two-)product commutators. (This expansion is presented in Eq (\ref{reduced}) to first order in the non-commutativity parameter.) We can map the star-commutators to matrix commutators and by expanding out these commutators, we realize that the three-bracket we have is nothing but the standard ABJM three-bracket of matrices (that is, on the form of Eq (\ref{abstract})) and of course this three-bracket satisfies the hermitian fundamental identity. It is not totally antisymmetric from a two-dimensional view-point where the three-algebra elements are functions $\t \T^a$ living on a two-dimensional space (or isomorphically matrices $\t T^a$). The remarkable thing is that the same three-bracket is totally antisymmetric from the three-dimensional view-point where the elements are on the form of Eq (\ref{red}), $\T^a = e^{i\sigma^3} \t \T^a$. If we restrict ourselves to such three-algebra generators only, then we can use another more useful inner product which then is also invariant. This inner product is presented below in Eq (\ref{inner}).

If on the other hand we allow for more general functions, say $\T^{a,p} = e^{ip\sigma^3} \t \T^a$ for some unspecified integers $p$, then we do not recover the standard ABJM three-bracket in two-dimensions (unless all the $p$'s are the same), and the fundamental identity is lost. Moreover, the inner product (\ref{inner}) is no longer invariant. The truncated and totally antisymmetric three-bracket does in general not satisfy the hermitian fundamental identity.

\subsection{On the 3/2-scaling}
For Chern-Simons levels $K=1,2$ the three-bracket becomes totally antisymmetric by means of monopole operators which give rise to certain identities \cite{Gustavsson:2009pm}. It was also speculated in \cite{Gustavsson:2009pm} that these identities which give rise to the totally antisymmetric three-bracket, could reduce the degrees of freedom to give us the 3/2-scaling behavior. One may now speculate that for $K=1,2$, we may still use the truncated three-bracket since this is indeed totally antisymmetric. One may speculate that truncation corresponds to taking into account the effect of monopole operators. If that is true, then we may think on the inner product as being on the form 
\ben
\<\T^a_A,\T^b_B\> &=& \int d^3 \sigma \sqrt{g} (\T^a_A \cdot 1)(\T_b^B \cdot 1).\label{inner}
\een
where the notation $\T^a_A \cdot 1$ can be thought of as the action of the three-algebra generator $T^a_A$ on the constant function which is equal to $1$ everywhere. In other words, $\T^a_A \cdot 1 = \T^a_0$. We see that the inner product only depends on equivalence classes $[\T^a_A]$ where we identify any two generators $\T^a_A \sim \T^b_B$ if $a=b$. This defines a gauge equivalence since the variation within the equivalence class does not affect the inner product. This inner product is not invariant, at least not in any obvious way and at a classical level. This is due to extra phase factors that comes in as a consequence of the truncated outer derivatives. It is still true that the magnitudes of $\<[\T^m_M,\T^p_P;\T^q_Q],\T^n_N\>$ and $\<\T^m_M,[\T^n_N,\T^q_Q;\T^p_P]\>$ as calculated using truncated inner product, agree. But they differ by their different phase factors. 

If we ignore this problem of understanding the trace invariance, and just use this truncated inner product, then we may obtain the 3/2-scaling of number of degrees of freedom. The dimension of the moduli space corresponds to the number of solutions to the equation $\<[\T^m_M,\T^n_N;\T^p_P],[\T^m_M,\T^n_N;\T^p_P]\> = 0$ (which in turn comes from the sextic potential in ABJM/BLG theory). This equation has a maximal set of solutions given by $\T^m_M$ with $m_3=0$ but with no restriction on $M$. We may refer to such a maximal set as a maximal set of three-commuting generators, or as a Cartan sub-three-algebra. The dimension of the Cartan sub-three-algebra is $N^2$ (times $N^3$ from the $M$-index). The total number of generators is $N^3$ (again times $N^3$ from the $M$-index). The generator $\T^m_M$ is gauge equivalent with $\T^m_0$ for $M=0$. We now get the $\N^{3/2}$ scaling of the number of gauge inequivalent generators, where $\N = N^2$ is the dimension of gauge inequivalent vacuum configurations (i.e. the Cartan modulo gauge redundances).

The usual ABJM theory whose three-algebra is realized by matrices, can be obtained from ABJM theory whose three-brackets are realized by star-three-products\footnote{It is possible that also star-three-products can be mapped into matrix multiplications, but we will not attempt to show this in this paper.}. To this end we consider the limit $K$ large. Here $K$ is an orbifolding of the fiber-direction, when the three-manifold is viewed as a fiber bundle. In our three-torus example, if we choose $\sigma^3$ as the fiber direction, the orbifold identification will be
\bea
\sigma^3 &\sim & \sigma^3 + \frac{2\pi}{K}.
\eea
In our finite truncation, we considered harmonics $\T^m = e^{im_{\aalpha}\sigma^{\aalpha}}$. Orbifolding restricts the possible set to those for which  
\ben
m_3 &=& Km_3 + 1, \label{nakwoo}
\een
($m\in \mb{Z}$) which are those harmonics which obey the Bloch wave condition on the orbifolded three-torus \cite{Kim:2008gn},
\bea
\T^m(\sigma^1,\sigma^2,\sigma^3 + 2\pi/K) &=& e^{\frac{i 2\pi}{K}} \T^m(\sigma^1,\sigma^2,\sigma^3).
\eea
Hence while $m_1,m_2 = 0,...,N-1$, we find that $m_3=0,...,[\frac{N-1}{K}]$ where $[\bullet]$ denotes the integer part. So when $K=N$ or larger, we must choose $m_3 = 1$. For those 'large' values of $K=N,N+1,...$ the star-three-product reduces to a star-product on the base-manifold (the two-torus spanned by $\sigma^{1}$ and $\sigma^2$). This implies that the star-three-product theory becomes equivalent to ABJM theory when $K=N,N+1,...$. 

Let us now compute the dimension of the Cartan $\N$ and the number of three-algebra generators $\D$ for generic $K$ and $N$. Let us ignore the $M^{\alpha}$ indices from now on, these being just a gauge redundancy. Since $\partial_3 \T^m \neq 0$ when orbifolding, we must choose Cartan generators with $m_3 \neq 0$. Let us choose the Cartan generators such that $m_{\aalpha} = (m,0,Kn+1)$. Then we obtain \cite{Gustavsson:2010nc} 
\bea
\N &=& N\([\frac{N-1}{K}] + 1\),\cr
\D &=& N^2\([\frac{N-1}{K}] + 1\).
\eea
For large $N$ and large $K$ we then get
\bea
\D &=& K^{1/2}\N^{3/2}
\eea
which agrees with the result in \cite{Drukker:2010nc} for large ´t Hooft coupling $\lambda = \N/K$. When $\lambda < 1$ or in other words when $K=N$ or larger, the above formula breaks down and we get instead
\bea
\D &=& \N^2.
\eea
In \cite{Drukker:2010nc} the number $\N$ is the number that appears in the gauge group of ABJM theory as $U(\N)\times U(\N)$. But also $\N$ in that approach coincides again with the dimension of the moduli space. This suggests that star-product theory which is characterized by an integer $N$, is the quantum effective theory of ABJM theory with gauge group characterized by the integer $\N$. In usual matrix realization ABJM theory we consider fixed gauge group $U(\N)\times U(\N)$ and vary the Chern-Simons level $K$ say. The common wisdom is that $\N$ counts the number of M2 branes, and we agree on that. However in star-three-product realization of ABJM theory it may appear more natural to keep $N$ fixed and vary $K$ say. But it is then that we get a quadratic type of behavior on the number of M2 branes as a function of $N$ for $K=1,...,N-1$. On the other hand, for $K=N,N+1,...$ the integer numbers $N$ and $\N$ coincide, and both count the number of M2 branes.

\subsection{Tensor product of three-algebras}\label{tensor}
In order to deconstruct non-Abelian D4 and M5 brane theories, we need to consider a tensor product of two three-algebras. We thus consider a tensor product $\A \otimes \B$ of two three-algebras $\A$ and $\B$. The generators of $\A\otimes \B$ are
\bea
T^{aa'} &=& T^a \otimes T^{a'}
\eea
where $\T^a \in \A$ and $\T^{a'} \in \B$. By repeated use of the abstractly defined three-bracket as in (\ref{abstract}) where the generators are three-multiplied by some (yet unspecified) three-multiplication, we can express the tensor-product three-bracket in the form
\ben
[T^{aa'},T^{bb'};T^{cc'}] &=& [T^a,T^b;T^c] \otimes T^{a'} T_{c'} T^{b'} + T^b T_c T^a \otimes [T^{a'},T^{b'};T^{c'}].\label{tensor1}
\een
and we may further expand the three-bracket as
\ben
[T^{a'},T^{b'};T^{c'}] &=& [T^{a'},T_{c'}T^{b'}] - [T^{b'},T_{c'}]T^{a'}.\label{tensor2}
\een
This expression is still defined by use of three-multiplication among the three elements involved. So for instance a commutator of just generators would have been ill-defined, but here we have a multiplication by a third element so we may use three-multiplication to carry out these products. 

One class of hermitian three-algebras has associated Lie algebras $U(N)\times U(N)$. These three-algebra generators also generate $U(N)$ Lie algebra \cite{Antonyan:2008jf}. The main example is $U(2)\times U(2)$ which has the same three-algebra as $SU(2)\times SU(2)$. As the associated three-algebra generators $T^a = (\sigma^i,i)$ generate $U(2)$ Lie algebra, their tensor products generate $U(4)$ Lie algebra. We now ask what is the corresponding three-algebra that these three-algebra generators generate? The answer is provided by \cite{Antonyan:2008jf}. The three-algebra is the one that comes with the associated Lie algebra $U(4)\times U(4)$. In general when we take the tensor product of $U(N_{\A})\times U(N_{\A})$ and $U(N_{\B})\times U(N_{\B})$ three-algebras, we get $U(N_{\A}N_{\B})\times U(N_{\A}N_{\B})$ three-algebra. A similar result appears to have been reached in \cite{Palmkvist}. It will be interesting to extend this to tensor products of any two ABJM gauge groups, and to understand what Lie algebra the three-algebra generate.

\subsection{Three-algebra of a three-sphere}
An example of a real three-algebra with complex generators is $SO(4)$ with three-algebra generators $T^i = (\frac{1}{2}\sigma^I, \frac{i}{2}\mb{I})$ where $\sigma^I\sigma^J = \delta^{IJ} + i\epsilon^{IJK}\sigma^K$. These generate the real three-algebra
\bea
[T^i,T^j;T^k] &=& -\frac{1}{2}\epsilon^{ijkl} T^l.
\eea
There is no geometrical interpretation of the matrix realization of this algebra. If we realize the same algebra by the Nambu bracket on $S^3$, then the $\T^i$ are real-valued coordinates in $\mb{R}^4$ describing the embedding of a round $S^3$. But this stands in conflict with the fact that $T^i$ as realized by matrices are not all hermitian. One may double the size of the matrices and consider hermitian $4\times 4$ gamma matrices $\gamma^i$ whose off-diagonal blocks are $T^i$ and $T_i$ respectively. But these doubled matrices do not close into an algebra, but we rather get something like $[\gamma^i,\gamma^j;\gamma^k] \sim \epsilon^{ijkl}\gamma_5 \gamma^l$. The presence of $\gamma_5$ means that this is not a closed algebra over the real (or complex) numbers.

These troubles may be an indication that we shall perhaps not try to realize $S^3$ by matrices at all. Matrices are excellent tools to realize two-manifolds \cite{Madore:1991bw}, but they do not seem to be suited for realizing three-manifolds. We may use matrices to realize a two-dimensional base-manifold of a three-dimensional fiber-bundle, but it seems we can not realize the whole three-manifold by matrices. 

Let us illustrate further how this works in our $SO(4)$ example. Let us define
\bea
G^2 &=& T^1 + iT^2,\cr
G^1 &=& T^3 + iT^4.
\eea
These $G^a$ are nothing but the GRVV matrices \cite{Gomis:2008vc}, \cite{Terashima:2008sy} for rank $N=2$. But instead of four independent hermitian matrices, we just have three independent since $G^1$ is hermitian, so we only have $G^1$, $G^2$ and $G_2$ as independent. We have the three-algebra (GRVV algebra)
\bea
[G^a,G^b;G^c] &=& -2\delta^{ab}_{cd} G^d
\eea
But since the matrix realization of this algebra only has three independent real generators it misses out some part of the $S^3$ geometry. In fact, and as we will explain shortly, the matrices only realize the fuzzy $S^2$ base-manifold of the $S^3$ viewed as a circle bundle \cite{Nastase:2010uy}.

If we expand the three-bracket as (\ref{surprise}) we see that the GRVV algebra reduces to an oscillator algebra,
\bea
[G^a,G^b] &=& 0,\cr
[G^a,G_b] &=& \delta^a_b
\eea
This in turn implies that we have an $SU(2)$ algebra induced from the GRVV algebra, generated by 
\bea
J_I &=& G^a (\sigma_I)_a{}^b G_b.
\eea
However the oscillator algebra only has the infinite-dimensional matrix realization. For a single oscillator this matrix is $(G^1)_{mn} = \sqrt{n} \delta_{m,n-1}$. 

The GRVV algebra also has finite-dimensional matrix realizations. For generic $N$, the GRVV matrices, as obtained independently by the authors of \cite{Gomis:2008vc} and \cite{Terashima:2008sy} respectively, and which were further studied in \cite{Kim:2010mr}, are given by
\bea
G^1 &=& \(\begin{array}{cccc}
0 & 0 & \cdots & 0\\
0 & \sqrt{1} & \cdots & 0\\
  &      &   \ddots & \\
0 &  0   &   \cdots       & \sqrt{N-1}
\end{array}\),\cr
G^2 &=& \(\begin{array}{ccccc}
0 & \sqrt{N-1} & \cdots & 0 & 0\\
0 & 0 & \cdots  & 0 & 0\\
  &   & \ddots &  & \\
0 & 0 & \cdots & 0 & \sqrt{1}\\
0 & 0 & \cdots & 0 & 0.
\end{array}\)
\eea
Again $G^1$ is hermitian. We also have the radius constraints
\bea
G^a G_a &=& (N-1)\mb{I},\cr
G_a G^a &=& N\(\mb{I}-E\)
\eea
where $\mb{I} = $diag$(1,1,...,1)$ and $E = $diag$ (1,0,...,0)$. The $G^a$ realize the above $SO(4)$ three-algebra for any $N$. 

Let us consider four real coordinates $\T^i$ in euclidean $\mb{R}^4$ describing the embedding of a round $S^3$. These are thus subject to the $SO(4)$ three-algebra and the radius constraint, 
\bea
\{\T^i,\T^j,\T^k\} &=& \frac{1}{R}\epsilon_{ijkl}\T^l,\cr
\T^i \T^i &=& R^2
\eea
To make the relation with the above $SO(4)$ algebra transparent, we define complex coordinates 
\bea
\G^1 &=& \T^1 + i \T^2,\cr
\G^2 &=& \T^3 + i \T^4
\eea
in terms of which we have
\bea
\{\G^a,\G^b,\G_c\} &=& \frac{4}{R}\delta^{ab}_{cd} \G^d,\cr
\G^a \G_a &=& R^2
\eea
In order to relate the functions $\G^a$ with the matrices $G^a$, we define a three-bracket 
\bea
[\G^a,\G^b;\G^c] &=& \hbar \{\G^a,\G^b,\G_c\}.
\eea
Then 
\bea
[\G^a,\G^b;\G^c] &=& \frac{4\hbar}{R}\delta^{ab}_{cd} \G^d,\cr
\G^a \G_a &=& R^2
\eea
We will keep $R$ fixed.

If we define an equivalence
\bea
\G^a &\sim & e^{i\psi} \G^a
\eea
then we have the isomorphism between such equivalence classes and matrices,
\bea
[\G^a] &\cong & \frac{R}{\sqrt{N-1}} G^a
\eea
if we take
\ben
\hbar &=& -\frac{R^3}{2(N-1)}.\label{hbar}
\een
We see that $\hbar\rightarrow 0$ as $N\rightarrow \infty$. We get a classical sphere in the large $N$ limit as expected. The reason we must consider equivalence classes is that the three-algebra is invariant under $\G^a \rightarrow e^{i\psi}\G^a$. 

We may parametrize the embedding coordinates $\G^a$ as
\ben
\G^1 &=& \frac{R}{\sqrt{2}} \sqrt{1+\cos\theta}e^{i(\varphi+\psi)},\cr
\G^2 &=& \frac{R}{\sqrt{2}} \sqrt{1-\cos\theta}e^{i\psi}\label{g}
\een
This parametrization makes the fiber-bundle structure manifest. It also enable a simple description of $S^3/{\mb{Z}_K}$. We just make the identification $\psi \sim \psi + \frac{2\pi}{K}$. The length of the fiber is $2\pi R/K$. Locally there is no difference between $S^3$ and $S^3/{\mb{Z}_K}$. We will define
\bea
\G^a &=& e^{i\psi} \t \G^a.
\eea
Then $\t \G^a$ is a representative in $[\G^a]$.

Let us define
\bea
g^a &=& \frac{R}{\sqrt{N-1}} G^a
\eea
We then have the isomorphism
\bea
g^a &\cong & \t G^a.
\eea
If we normalize the trace form as $\<g^a,g^b\> = \<\t \G^a,\t \G^b\> = \delta^a_b$, we have the realizations
\bea
\<g^a,g^b\> &=& \frac{2}{R^2 N} \tr(g^a g_b),\cr
\<\t \G^a, \t \G^b\> &=& \frac{1}{2\pi R^2} \int d\theta d\varphi \sin \theta \t \G^a \t \G_b.
\eea
and we can make the identification
\ben
\frac{1}{N} \tr &\cong & \frac{1}{\pi R^2} \int d\theta d\varphi \(\frac{R}{2}\)^2\sin \theta.\label{norms}
\een

If we define the Hopf projection
\bea
X^I &=& \G^a (\sigma^I)_a{}^b \G_b
\eea
where $\sigma^I$ are the Pauli matrices, or explicitly 
\bea
X^1 &=& \G^1 \G_2 + \G^2 \G_1,\cr
X^2 &=& -i\G^1 \G_2 + i\G^2\G_1,\cr
X^3 &=& \G^1\G_1 - \G^2\G_2
\eea
then we get
\bea
X^1 &=& R^2 \sin \theta \cos \varphi,\cr
X^2 &=& -R^2 \sin \theta \sin \varphi,\cr
X^3 &=& R^2 \cos \theta.
\eea
By using the Fierz identity
\bea
(\sigma^I)_a{}^b (\sigma^I)_c{}^d &=& 2\delta_a^d \delta_b^c - \delta_a^b \delta_c^d
\eea
we find that
\bea
\frac{1}{4R^2} dX^I dX^I &=& d\G^a d\G_a - R^2 (d\psi + A_{\varphi} d\varphi)^2
\eea
More explicitly we get
\bea
\frac{1}{4R^2} dX^I dX^I &=& \frac{R^2}{4}\(d\theta^2 + \sin^2\theta d\varphi^2\),\cr
d\G^a d \G_a &=& \frac{R^2}{4} \(d\theta^2 + \sin^2 \theta d\varphi^2\) + R^2 \(d\psi + A_{\varphi} d\varphi\)^2
\eea
where
\bea
A_{\varphi} &=& \frac{1}{2}\(1+\cos\theta\)
\eea
is the gauge field of a magnetic monopole of unit one. Its field strength is
\bea
F_{\varphi\theta} &=& \frac{1}{2}\sin\theta
\eea
and the integral over $S^2$ is 
\bea
\int d\varphi d\theta F_{\varphi\theta} &=& 2\pi.
\eea
We will denote the coordinates by $\sigma^{\aalpha} = (\sigma^m,\psi) = (\theta,\varphi,\psi)$ and the metric tensors by $G_{mn}$ on $S^2$ and $g_{\aalpha\bbeta}$ on $S^3$. Their square root determinants are given by  
\ben
\sqrt{g} &=& \frac{R^3}{4}\sin\theta,\cr
\sqrt{G} &=& \(\frac{R}{2}\)^2\sin\theta.\label{metrics}
\een
The radius of $S^2$ is $\frac{R}{2}$, the radius of the fiber is $R$ which is the same as the radius of $S^3$. This can also be confirmed by a computatation of the volume of $S^3$,
\bea
\int_0^{2\pi} d\psi \int_0^{\pi} d\theta \int_0^{2\pi} d\varphi \frac{R^3}{4}\sin\theta &=& (2\pi R) \(4\pi \(\frac{R}{2}\)^2\)
\eea
The result is $2\pi^2 R^3$ as is the volume of $S^3$ with radius $R$, but it can be computed as the length of the fiber which is $2\pi R$ irrespectively of at which point on the base-manifold it is evaluated, times the volume of the base-manifold, which equals the volume of $S^2$ of radius $\frac{R}{2}$.

\subsection{Three-algebra basis on a fuzzy three-sphere}
We would like to consider the algebra of functions on $S^3$. If we consider the infinite set of functions we can three-multiply together $\G^a$ and $\G_a$ by using the usual multiplication of two functions iteratively. If we want to consider a finite truncation we must use star-three-multiplication \cite{Gustavsson:2010nc}. The important properties are associativity and that upon dimensional reduction it is mapped isomorphically into matrix multiplication. We may then consider functions
\bea
\T^{\vec{a}} &=& (\G^{a_1} * \G_{b_1} * \G^{a_2} *  ... * \G^{a_{r-1}} * \G_{b_{r-1}} * \G^{a_{r}}) \cdot 1
\eea
These all have a trivial dependence on the fiber given by the phase factor $e^{i\psi}$. We may write
\bea
\T^{\vec{a}}(\theta,\varphi,\psi) &=& e^{i\psi} \t \T^{\vec{a}}(\theta,\varphi).
\eea
Since these functions all have the same trivial dependence on the fiber, it is clear that these do not constitute a basis of all functions on $S^3$. In fact, these $\T^a$ already have the dimensional reduced form (\ref{red}) and shall be associated with the $S^2$ base manifold. Since we thus consider both truncation and dimensional reduction, we may replace all star-three-products with two-dimensional star products for free. We may then turn to the isomorphic matrix realization where we have the basis elements
\bea
T^{\vec{a}} &=& G^{a_1} G_{b_1} G^{a_2} ... G^{a_{r-1}} G_{b_{r-1}} G^{a_{r}}
\eea
for $r=1,...,N-1$. This set is finite so we may count how many they are. We may express any element generated by this basis as a linear combination 
\ben
M &=& c_a G^a + c_{a_1 a_2}^{b_1} G^{a_1} G_{b_1} G^{a_2} + \cdots + c_{a_1\cdots a_{\l}}^{b_1 \cdots b_{\l-1}} G^{a_1}G_{b_1}\cdots G_{b_{\l-1}} G^{a_{\l}}\cr
&& + \cdots + c_{a_1\cdots a_{N-1}}^{b_1 \cdots b_{N-2}} G^{a_1}G_{b_1}\cdots G_{b_{N-2}} G^{a_{N-1}}.\label{expansion}
\een
We ask how many independent components $c_{a_1\cdots a_{\l}}^{b_1 \cdots b_{\l-1}}$ we have. For commuting functions we find symmetrized indices $a_1\cdots a_{\l}$ as well as $b_1 \cdots b_{\l-1}$. Hence we have $(\l+1)\l$ components. But these are not all independent because of the sphere constraint $G^a G_a = R^2$. For matrices we only need to consider down traces since up traces are related to down traces by the sphere equation. For commuting functions down and up traces are the same, by just commuting the functions. We then need to remove the number of components in a down trace say. There are $\l(\l-1)$ such components. We are left with $2\l$ independent components in $c_{a_1\cdots a_{\l}}^{b_1 \cdots b_{\l-1}}$. Summing them up we get 
\bea
\sum_{\l = 1}^{N-1} 2\l &=& (N-1)N
\eea
components in total. We have $N\times (N-1)$ matrices, and indeed we may truncate the GRVV matrices to size $N\times (N-1)$ and then their conjugates will be of size $(N-1)\times N$. Then it is easy to see that ${T}^{\vec{a}}$ is an $N\times (N-1)$ matrix, and any $N\times (N-1)$ matrix can be expressed as a linear combination $c_{\vec{a}} T^{\vec{a}}$. The GRVV matrices have real entries (though they are not hermitian). We may then restrict ourselves to real coefficients $c_{\vec{a}}$ and to $N\times (N-1)$ matrices with real entries. It is now clear that $T_{\vec{a}}$ (which equals the transpose of $T^{\vec{a}}$), is an $(N-1)\times N$ matrix. Ignoring the mismatch of $\O(1/N)$, we may expand this matrix in the basis of $N\times (N-1)$ matrices as
\bea
T_{\vec{a}} &=& \kappa_{\vec{a}\vec{b}} T^{\vec{b}}.
\eea
Clearly $T^{\vec{a}}$ must generate the three-algebra which is associated to $U(N)\times U(N)$ gauge group (modulo the $\O(1/N)$ mismatch).

We have a basis of functions on a fuzzy $S^2$ (ignoring $1/N$ mismatch) corresponding to spherical harmonics with $\ell = 0,...,N-1$. A basis for functions on the fuzzy $S^3$ is 
\bea
\T_m^{\vec{a}} &=& e^{im\psi} \t {\T}^{\vec{a}}
\eea
where $m$ ranges over $[-N-1,N-1]$, that is $m$ takes $2N-1$ different values. In particular when $N=2$ we have the basis elements $e^{\pm i\psi} \t \G^a$ and $e^{\pm i \psi} \t \G_a$ which correspond to the complete set of embedding coordinates $\G^a$ and $\G_a$ of $S^3$ in $\mb{R}^4$. The number of M2 branes should be given by the dimension of the moduli space. These are the three-commuting functions. One maximal set of three-commuting functions is given by $\t {\T}^{\vec{a}}$. Since these only depend on two coordinates, these have vanishing three-bracket (and in particular vanishing Nambu bracket). The number of degrees of freedom should on the other hand be proportional to the dimension of the span of ${\T}^{\vec{a}}_m$ where the matter fields are valued. The former is $\N = N(N-1)$, the latter is $\D = N(N-1)(2N-1)$. Orbifolding the three-sphere restricts $m$ to take $[(2N-1)/K]$ different values. When $K = 2N$ or larger, we find $\D \propto \N^2$.

\section{The D4 brane}\label{The D4 brane}
The theory of multiple M5's should reduce to the theory of multiple D4's upon compactification on a circle. We may also derive D4 from M2 by fluctuation analysis about a nontrivial vacuum solution. By so identifying these two D4 brane theories, we will be able to derive the M5 coupling constant (which is uniquely fixed) directly from the M2 brane theory. 

\subsection{Single M5 reduced to single D4}
Let us start by a single M5 which has $(2,0)$ supersymmetry in six dimensions. We assume eleven-dimensional gamma matrices $\Gamma_M$ and $\Gamma_I$ where $M=0,1,2,3,4,5$ and $I=6,7,8,9,10$. We define $\Gamma = \Gamma_{012345}$ and assume a supersymmetry parameter subject to 
\bea
\Gamma \omega &=& \omega
\eea
The fermions in $(2,0)$ tensor multiplet are now subject to 
\bea
\Gamma \chi &=& -\chi
\eea
We have the supersymmetry variations
\bea
\delta \phi^I &=& i\bar{\omega}\Gamma^I\chi,\cr
\delta B_{MN} &=& i\bar{\omega}\Gamma_{MN}\chi,\cr
\delta \chi &=& \frac{1}{12}\Gamma^{MNP}\omega H_{MNP} + \Gamma^M \Gamma_I \omega \partial_M \phi^I
\eea
and reduce on a circle, which means split $M=(\mu,\psi)$ and ignore derivatives with respect to $\psi$. We then get (for a precise definition of the dimensionally reduced fields, we refer to eq (\ref{assign}))
\bea
\delta \phi^I &=& i\bar{\omega}\Gamma^I\chi,\cr
\delta A_{\mu} &=& i\bar{\omega}\Gamma_{\mu}\Gamma_{\psi}\chi,\cr
\delta \chi &=& \Gamma^{\mu}\Gamma_I\omega \partial_{\mu}\phi^I+\frac{1}{2}\Gamma^{\mu\nu}\Gamma^{\psi}\omega F_{\mu\nu}.
\eea
but still we have the eleven-dimensional spinor quantities which are subject to unusual chirality conditions. From ten-dimensional point, $\Gamma_{\psi}$ is the natural chirality matrix with respect to which we shall define chiralities of our spinors. Hence we like to use supersymmetry parameter $\epsilon$ and spinor field $\psi$ subject to chirality conditions
\bea
\Gamma_{\psi}\epsilon &=& \epsilon,\cr
\Gamma_{\psi}\psi &=& -\psi.
\eea
Such spinors can be related to the previous ones by a unitary rotation
\bea
\omega &=& U \epsilon,\cr
\chi &=& U \psi
\eea
where
\bea
U &=& \frac{1}{\sqrt{2}} (1 + \sigma),\cr
\sigma &=& \Gamma_{01234}.
\eea
Here we have the property $\sigma^2 = -1$ and $\sigma^{\dag} = -\sigma$. If we also make a field redefinition 
\bea
\xi &=& \sigma \psi
\eea
which changes the chirality condition as
\bea
\Gamma_{\psi} \xi &=& \xi
\eea
then in terms of these redefined spinors we have
\bea
\delta \phi^I &=& i\bar{\epsilon}\Gamma^I \xi,\cr
\delta A_{\mu} &=& i\bar{\epsilon}\Gamma_{\mu}\xi,\cr
\delta \xi &=& \frac{1}{2}\Gamma^{\mu\nu}\epsilon F_{\mu\nu}+\Gamma^{\mu}\Gamma_I\epsilon \partial_{\mu}\phi^I.
\eea
Here we have be able to completely eliminate all $\Gamma^{\psi}$ using the chirality condition of $\xi$. Still we work with gamma matrices that anti-commute
\bea
\{\Gamma^I,\Gamma_{\mu}\} &=& 0.
\eea

\subsection{Multiple D4}
Now we have obtained exactly the supersymmetry variations that we would derive if we reduce ten-dimensional Abelian super Yang-Mills to five dimensions. But of course it would be no more difficult to reduce non-Abelian super Yang-Mills. If we do, then we get the supersymmetry variations
\bea
\delta \phi^I &=& i\bar{\epsilon}\Gamma^I \xi,\cr
\delta A_{\mu} &=& i\bar{\epsilon}\Gamma_{\mu}\xi,\cr
\delta \xi &=& \frac{1}{2}\Gamma^{\mu\nu}\epsilon F_{\mu\nu}+\Gamma^{\mu}\Gamma_I\epsilon D_{\mu}\phi^I + \frac{1}{2}\epsilon[\phi^I,\phi^I].
\eea
The challenge now is to see whether these variations can be derived from a proposed non-Abelian $(2,0)$ theory.

The super Yang-Mills Lagrangian is given by
\bea
\frac{1}{g_{YM}^2} \tr\(-\frac{1}{4}F^{\mu\nu}F_{\mu\nu}-\frac{1}{2}D^{\mu}\phi^I D_{\mu}\phi^I + \frac{1}{4}[\phi^I,\phi^J]^2\).
\eea
where 
\bea
D_{\mu}\phi^I &=& \partial_{\mu}\phi^I + i[A_{\mu},\phi^I],\cr
F_{\mu\nu} &=& \partial_{\mu}A_{\nu} - \partial_{\nu}A_{\mu} + i[A_{\mu},A_{\nu}]
\eea
Alternatively we have
\bea
\tr\(-\frac{1}{4}F^{\mu\nu}F_{\mu\nu}-\frac{1}{2}D^{\mu}\phi^I D_{\mu}\phi^I + \frac{g^2}{4}[\phi^I,\phi^J]^2\).
\eea
where 
\bea
D_{\mu}\phi^I &=& \partial_{\mu}\phi^I + ig_{YM}[A_{\mu},\phi^I],\cr
F_{\mu\nu} &=& \partial_{\mu}A_{\nu} - \partial_{\nu}A_{\mu} + ig_{YM}[A_{\mu},A_{\nu}]
\eea
These two formulations are related by the field rescaling
\bea
\phi^I &\rightarrow & g_{YM} \phi^I,\cr
A_{\mu} &\rightarrow & g_{YM} A_{\mu}.
\eea

\subsection{The Yang-Mills coupling constant from M5}
If we start with M five-brane action \cite{Henningson:2004dh}\footnote{Here the spacetime index runs over $M=0,1,...,5$ and $H_{MNP}$ is not selfdual. However the antiselfdual part decouples and is not part of the $(2,0)$ tensor multiplet and does not occur in the supersymmetry variations. The decoupling of the antiselfdual piece of $H$ gets clear when we couple this action to a background $C$ field as in \cite{Witten:1996hc} where only the selfdual piece of $H$ couples to $C$. The decoupling of the antiselfdual piece is best understood by carrying out a holomorphic factorization of the partition function \cite{Witten:1996hc}, \cite{Henningson:1999dm}. Of course the partition function for the M five-brane is just a holomorphic factor, and the action for the nonselfdual two form is a tool to obtain the M five-brane partition function via holomorphic factorization. Keeping holomorphic factorization in mind, it seems fair to say that this action can be used to describe a single M five brane.}
\bea
\frac{1}{4\pi}\int d^6 x\(-\frac{1}{12}H^{MNP}H_{MNP} - \partial_M \phi^I \partial^M \phi^I\)
\eea
and dimensionally reduce it on a circle with circumference $2\pi R$, define new dimensionally reduced fields as 
\ben
A_{\mu} &=& 2\pi RB_{\mu\psi},\cr
\phi^I &=& 2\pi R\phi^I\label{assign}
\een
then we get
\bea
\frac{1}{4\pi^2 R}\int d^5 x \(-\frac{1}{4}F^{\mu\nu} F_{\mu\nu} - \frac{1}{2}\partial_{\mu}\phi^I \partial^{\mu}\phi^I\).
\eea
We read off the Yang-Mills coupling constant
\bea
g^2_{YM} &=& 4 \pi^2 R.
\eea
The coupling becomes infinite as $R\rightarrow \infty$. It has been argued that the strong coupling limit of five-dimensional super Yang-Mills might be six-dimensional $(2,0)$ theory \cite{Tong:2005un}, \cite{Lambert:2010iw}, \cite{Douglas:2010iu}.

Of course in this example the theory is free and we may not interpret $g_{YM}$ as a coupling constant. But we can get non-Abelian super Yang-Mills theory essentially by replacing $\partial_{\mu}$ by $\partial_{\mu} + [A_{\mu},\bullet]$ (and adding some more interaction terms). Doing that, we conclude that $g_{YM}$ is the coupling constant. Also it is important to note that the Dirac charge quantization condition is inherited from six dimensions as
\bea
\int d\psi \wedge dX^{\mu} \wedge dX^{\nu} H_{\mu\nu\psi} = \int dX^{\mu} \wedge dX^{\nu} F_{\mu\nu}
\eea
as\footnote{Here we work in a convention where $\psi$ has mass dimension $-1$ so that $H_{\mu\nu\psi}$ will have the usual engineering mass dimension $3$.} $\psi \in [0,2\pi R]$  and the $X^{\mu}$ are coordinates describing the embedding of some two-manifold in five dimensions.

\section{Star-three-product BLG theory}\label{ABJM theory}
It was noted in \cite{Kim:2008gn}, that BLG theory\footnote{Usually BLG theory is thought of as ABJM theory with gauge group $SO(4)$. But BLG theory is more than that. We also have BLG theories for the Nambu bracket. Here we extend this considerably by using truncated star-three-product bracket in place of the Nambu bracket, which enable us to use BLG for finite-dimensional three-algebras whose generators are on the form of Eq (\ref{red}). These are not really new theories. They are the good old ABJM theories, though expressed from a three-dimensional vantage point. We could also just stick to the usual formulation of ABJM theory.} on an orbifold is equivalent with ABJM theory. Here we will extend that idea to our star-three-bracket. The star-three-bracket 
\bea
[\T^a,\T^b;\T^c] &=& \T^a * \T_c * \T^b - \T^b * \T_c * \T^a
\eea
is not totally antisymmetric. But the truncated three-bracket
\bea
[\T^a,\T^b,\T_c] &\equiv & [\T^a,\T^b;\T^c]\cdot 1
\eea
is indeed totally antisymmetric \cite{Gustavsson:2010nc}, and this we indicate by using comma to separate the three entries, instead of a semi-colon. In the process we must also change the third entry to its conjugate element. To leading order this totally antisymmetric bracket is equal to the Nambu bracket,
\bea
[\T^a,\T^b,\T_c] &=& \hbar \{\T^a,\T^b,\T_c\} + \O(\hbar^2).
\eea
Moreover, by restricting ourselves to generators of the form of Eq (\ref{red}), this totally antisymmetric three-bracket will satisfy the hermitian fundamental identity. Since the three-bracket is already truncated, we can safely use the inner product $(\ref{inner})$ and this will be invariant on the restricted set of generators which are on the form of Eq (\ref{red}). 

The total antisymmetry implies that the three-algebra is real, since we can not separate $\T_a$ from $\T^a$ as these are related by antisymmetry of the three-bracket. This justifies using the notation $[\T^a,\T^b,\T_c]$ for this real and totally antisymmetric three-bracket. Real in the sense that this bracket is complex conjugate linear in all three entries. It does not mean that all the $\T^a$ have to be real. It only means that $\T_a$ must be expressible as a linear combination of the $\T^a$, and we can surely find a linear combination of generators so they all become real. 

Now since three-brackets in this ABJM Lagrangian are totally antisymmetric (from the three-dimensional view-point), we can rewrite the ABJM Lagrangian as a BLG theory \cite{BLG}, \cite{G} with manifest $SO(8)$ symmetry. This is true for any Chern-Simons level $K$, and $SO(8)$ is only broken explicitly when $K>2$ by the orbifold identification $\mb{C}^4/{\mb{Z}_K}$. 

This means that supersymmetry enhancement in ABJM theory becomes a triviality when using the star-three-product formalism. If we use matrix realization of the gauge group $U(N)_K\times U(N)_{-K}$, we must use monopole operators to understand the supersymmetry enhancement for $K=1,2$ \cite{Bashkirov:2010kz}, \cite{Gustavsson:2009pm}.

There are some subtleties with applying the star-three-product formalism in ABJM and BLG theory though. First of all, we need a three-manifold on which we can define the star-three-product One way of generating such a three-manifold, is by mass deforming ABJM theory, so that it possesses a two-sphere vacuum solution. This two-sphere can then be interpreted as the base-manifold of $S^3/{\mb{Z}_K}$. But also, the identification of $K$ with the Chern-Simons level is not entirely obvious, even though it appears that Chern-Simons level $K$ corresponds M2 branes probing the $\mb{C}^4/{\mb{Z}_K}$ orbifold singularity \cite{ABJM}. The intersection of this orbifold and $S^3$ is of course the orbifolded $S^3/{\mb{Z}_K}$. The second subtlety is the isomorphism from functions being star-three-multiplied, to matrices and the gauge group $U(N)\times U(N)$. As we saw, we could only generate $N\times (N-1)$ bifundamental matrices by multiplying together an odd number of GRVV matrices, and maybe that should be taken seriously as saying that the gauge group we get from $S^3/{\mb{Z}_K}$ really corresponds to $U(N)\times U(N-1)$ rather than $U(N)\times U(N)$. This remains to be analyzed in more detail. In this paper we just ignore this discrepancy, and treat it like a $1/N$ correction. 

Eventhough we may define our star-three-product over $S^3$, we can not really probe the $S^3$ structure as we are confined to the subset of three-algebra generators which are on the form of Eq (\ref{red}). To really see the $S^3$ structure we would need to extend this to a complete set of functions on $S^3$, but this is not clear how to do and will presumably require the use of monopole operators. 

By using triality of $SO(8)$ in BLG theory we may take the eight scalar fields $X^{\alpha}$ to transform as an $SO(8)$ Weyl spinor, the eight spinors $\psi_{\dot{\alpha}}$ transform as a Weyl cospinor, and the eight supersymmetry parameters $\epsilon^I$ as vector. We assume a real three-algebra with totally antisymmetric three-bracket, but may use a complex basis for the three-algebra generators. We define the covariant derivative as
\bea
D_{\mu} X^{\alpha} &=& \partial_{\mu} X^{\alpha} + [X^{\alpha},T^c,T_d]A_{\mu}{}^d{}_c.
\eea
With these assignments, we have the $\N=8$ supersymmetry variations
\bea
\delta X^{\alpha} &=& -i\bar{\epsilon}_I\psi_{\dot{\beta}}\Gamma^{I\dot{\beta}\alpha},\cr
\delta \psi_{\dot{\alpha}} &=& \gamma^{\mu}\epsilon^I D_{\mu}X^{\alpha}\Gamma_{I\alpha\dot{\alpha}} -\frac{1}{6}\Gamma_{K\alpha\dot{\beta}}\Gamma^{J\dot{\beta}\gamma}\Gamma_{J\delta\dot{\alpha}}\epsilon^K [X^{\alpha},X^{\delta},X_{\gamma}],\cr
\delta A_{\mu} &=& i\bar{\epsilon}_I \gamma_{\mu} \Gamma^{I\dot{\alpha}\beta} [\bullet,\psi_{\dot{\alpha}},X_{\beta}].
\eea
Here 
\bea
\Gamma_I &=& \(\begin{array}{cc}
0 & \Gamma_{I\alpha\dot{\beta}}\\
\Gamma^{I\dot{\alpha}\beta} & 0
\end{array}\)
\eea
are $SO(8)$ gamma matrices, and $\gamma_{\mu}$ are $SO(1,2)$ gamma matrices. 

If we decompose the matter fields as
\bea
X^{\alpha} &=& \(\begin{array}{c}
Z^A\\
Z_A
\end{array}\)
\eea
and 
\bea
\psi_{\dot{\alpha}} &=& \(\begin{array}{c}
\psi^A\\
-\psi_A
\end{array}\).
\eea
and let $X_{\alpha} = (X^{\alpha})^*$ and $Z_A = (Z^A)^*$, one may check that the $\N=8$ supersymmetry variations
reduce to $\N=6$ variations of ABJM type
\bea
\delta Z^A &=& -i\bar{\epsilon}^{AB}\psi_B,\cr
\delta \psi_A &=& \gamma^{\mu}\epsilon_{AB}D_{\mu}Z^B - \epsilon_{AB}[Z^B,Z^C,Z_C] - \epsilon_{BC}[Z^B,Z^C,Z_A],\cr
\delta A_{\mu} &=& i\bar{\epsilon}_{AB}\gamma_{\mu}[\bullet,Z^A,\psi^B]-i\bar{\epsilon}^{AB}\gamma_{\mu}[\bullet,\psi_A,Z_B]
\eea
We have used scalar fields $X^{\alpha}$. Valued in three-algebra these have components
\bea
X^{\alpha} &=& \(\begin{array}{c}
Z^A_a T^a\\
Z_A^a T_a
\end{array}\).
\eea
If the three-algebra is real, we include $T_a = (T^a)^*$ as generators of the three-algebra. But we also have stressed that $T_a$ are not independent but can rather be expressed as a linear combination of $T^a$, 
\bea
T_a &=& \kappa_{ab} T^b.
\eea
We may then write the scalar fields in the form
\bea
X^{\alpha} &=& \(\begin{array}{c}
Z^A_a T^a\\
Z_{Aa} T^a
\end{array}\)
\eea
where $Z_{Aa} = \kappa_{ab} Z_A^b$. However this latter form is more subtle as it is no longer manifest in this notation that $Z_{Aa} T^a$ is the complex conjugate of $Z^A_a T^a$. This fact is very important as it means that we do not double the field content. The components of $X^{\alpha}$ are complex, but we do not double the number of independent components. We note that $\kappa_{ab}$ in star-three-product ABJM theory appears to play a role similar to that of a monopole operator in usual ABJM theory.

We may consider the triality map $(I,\alpha,\dot{\beta}) \mapsto (\dot{\alpha},I,\beta)$. This maps the two half-gamma matrices  
\bea
\Gamma_{I\alpha\dot{\beta}},\qquad \Gamma^{I\dot{\alpha}\beta}
\eea
into 
\bea
\Gamma_{I\beta\dot{\alpha}},\qquad \Gamma^{I\dot{\beta}\alpha}
\eea
if we define a cyclic symmetry $\Gamma_{I\alpha\dot{\beta}} = \Gamma_{\dot{\beta}I\alpha} = \Gamma_{\alpha\dot{\beta}I}$. We then find the supersymmetry variations
\bea
\delta X^I &=& -i\bar{\epsilon}_{\dot{\alpha}}\Gamma^{I\dot{\alpha}\beta}\psi_{\beta},\cr
\delta \psi_{\alpha} &=& \gamma^{\mu}\Gamma_{I\alpha\dot{\alpha}}\epsilon^{\dot{\alpha}}D_{\mu}X^I - \frac{1}{6} (\Gamma_K \Gamma^J \Gamma_I)_{\alpha\dot{\alpha}} \epsilon^{\dot{\alpha}}[X^I,X^K,X_J],\cr
\delta A_{\mu} &=& i\bar{\epsilon}_{\dot{\alpha}}\gamma_{\mu}\Gamma^{I\dot{\alpha}\alpha}[\bullet,\psi_{\alpha},X_I].
\eea

To compare with BLG theory as originally formulated in \cite{BLG}, we convert to eleven dimensional gamma matrices
\bea
\Gamma^{\mu} &=& \gamma^{\mu} \otimes \Gamma,\cr
\Gamma_I &=& 1 \otimes \Gamma_I
\eea
Along with this, we also declare that $X_I = X^I$. We assume chirality conditions
\bea
\Gamma \psi &=& -\psi,\cr
\Gamma \epsilon &=& \epsilon.
\eea
If we also shift the sign of the fermion $\psi \rightarrow -\psi$, then in terms of these gamma matrices the $\N=8$ supersymmetry variations read
\bea
\delta X^I &=& i\bar{\epsilon}\Gamma_I \psi,\cr
\delta \psi &=& \Gamma^{\mu} \Gamma_I \epsilon D_{\mu}X^I + \frac{1}{6}\Gamma_{IJK}[X^I,X^J,X^K],\cr
\delta A_{\mu} &=& i\bar{\epsilon}\Gamma^{\mu}\Gamma_I [\bullet,X_I,\psi].
\eea
where we have noted the identity
\bea
\Gamma_I \Gamma_J \Gamma_K - \Gamma_K \Gamma_J \Gamma_I &=& 2\Gamma_{IJK}.
\eea

The gauge variations are most clearly expressed in Lie algebra language as
\ben
\delta X^I &=& \Lambda(X^I),\label{deltaX}\\
\delta A_{\mu} &=& -D_{\mu} \Lambda.\label{deltaA}
\een
Here gauge covariant derivatives are defined as
\bea
D_{\mu} X^I &=& \partial_{\mu} X^I + A_{\mu}(X^I),\cr
D_{\mu} \Lambda &=& \partial_{\mu} \Lambda + [A_{\mu},\Lambda].
\eea
In three-algebra notation we have
\bea
\Lambda(X^I) &=& [X^I,T^c;T^d] \Lambda^d{}_c\cr
&=& \t\Lambda^b{}_a X^I_b T^a
\eea
where, if we let $[T^b,T^c;T^d] = f^{bc}{}_{da} T^a$, 
\bea
\t \Lambda^b{}_a &=& f^{bc}{}_{da} \Lambda^d{}_c.
\eea

\subsection{Mass deformation}
We can mass deform these variations and still keep the $\N=8$ supersymmetry while breaking $SO(8)$ down to $SO(4)\times SO(4)$ \cite{Gomis:2008cv}, \cite{Hosomichi:2008qk}. The mass deformation amounts to adding the term to the supersymmetry variation of the fermions
\bea
\delta \psi_{\alpha} &=& m  X^I{\Gamma_{(4)}}_{\alpha}{}^{\beta} \Gamma_{I\beta\dot{\gamma}} \epsilon^{\dot{\gamma}}
\eea
Here 
\bea
\Gamma_{(4)} &=& \Gamma_{1234}.
\eea
In the trial version this reads
\bea
\delta \psi_{\dot{\alpha}} &=& m X^{\alpha} \Gamma_{I\alpha\dot{\beta}} \Gamma_{(4)}{}^{\dot{\beta}}{}_{\dot{\alpha}} \epsilon^I.
\eea
and this we can also rewrite in the form of ABJM theory. 
\bea
\delta \psi_A &=& m \epsilon_{BC} G_A^B Z^C 
\eea
in ABJM theory. Here the matrix $G^C_B$ will be defined in eq's (\ref{G1}), (\ref{G2}).

\subsection{The mass deformed BLG Lagrangian}
The Chern-Simons term is unaffected by $SO(8)$ triality and will thus always look the same. It is given by
\ben
\L_{CS} &=& \frac{1}{2} \epsilon^{\mu\nu\lambda} \Big(\<T^b,[T^c,T^d;T^a]\> A_{\mu}{}^c{}_b \partial_{\nu} A_{\lambda}{}^d{}_a \cr
&& - \frac{2}{3} \<[T^a,T^c;T^d],[T^f,T^b;T^e]\>A_{\mu}{}^b{}_a A_{\nu}{}^d{}_c A_{\lambda}{}^f{}_e\Big).\label{CS}
\een
In trial version of general BLG theory, the matter Lagrangian and mass deformation terms are given by 
\bea
\L_{kin} + \L_V + \L_{Yukawa} &=& -\frac{1}{2}\<D_{\mu}X^{\alpha},D^{\mu}X^{\alpha}\>-\frac{1}{12}\<[X^{\alpha},X^{\beta};X^{\gamma}],[X^{\alpha},X^{\beta};X^{\gamma}]\>\cr
&&+\frac{i}{2}\<\bar{\psi}_{\dot{\alpha}},\gamma^{\mu}D_{\mu}\psi_{\dot{\alpha}}\> - \frac{i}{4} \<X^{\alpha},[X^{\beta},\bar{\psi}_{\dot{\alpha}};\psi_{\dot{\beta}}]\>\Gamma_{I\alpha\dot{\alpha}}\Gamma^{I\dot{\beta}\beta},
\eea
\bea
\L_{m} + \L_{flux} &=& -\frac{m^2}{2}\<X^{\alpha},X^{\alpha}\> + \frac{m}{48} \Gamma_{I\alpha\dot{\beta}} \Gamma^{I\dot{\epsilon}\beta}\Gamma_{K\gamma\dot{\epsilon}}\Gamma^{K\dot{\gamma}\delta}G^{\dot{\beta}}_{\dot{\gamma}} X^{\alpha} X^{\gamma} X_{\delta} X_{\beta}
\eea
where
\ben
G &=& \frac{1}{2}\(\Gamma_{1234} + \Gamma_{\hat{1}\hat{2}\hat{3}\hat{4}}\).\label{G1}
\een
In the original BLG theory we have
\bea
\L_{m} + \L_{flux} &=& -\frac{m^2}{2}\<X^I,X^I\>+\frac{m}{6}\(\epsilon_{ijkl}\<X^i,[X^j,X^k,X^l]\>+\epsilon_{\hat{i}\hat{j}\hat{k}\hat{l}}\<X^{\hat{i}},[X^{\hat{j}},X^{\hat{k}},X^{\hat{l}}]\>\).
\eea

\subsection{Rewriting in the form of ABJM}
We split $X^{\alpha}$ into Weyl components $Z^A$ and $Z_A$ and we get
\bea
\L_{kin} &=& -\frac{1}{2}\<D_{\mu}X^{\alpha},D^{\mu}X^{\alpha}\> \cr
&=& -\<D_{\mu} Z^A,D^{\mu} Z^A\>.
\eea
Using the fundamental identity we may also show that 
\bea
\L_V&=&-\frac{1}{12}\<[X^{\alpha},X^{\beta};X^{\gamma}],[X^{\alpha},X^{\beta};X^{\gamma}]\> \cr
&=& -\frac{2}{3}\(\<[Z^A,Z^B;Z^C],[Z^A,Z^B;Z^C]\> - \frac{1}{2}\<[Z^C,Z^A;Z^C],[Z^B,Z^A;Z^B]\>\).
\eea
By some work\footnote{We may realize the $SO(8)$ gamma matrices as $\Gamma^I = (\Sigma^M \otimes \sigma^1,1\otimes \sigma^2,\Sigma \otimes \sigma^1)$ for which we have the Fierz identities $(\Gamma^L_X)_A{}B (\Gamma^R_X)_C{}^D = 2\delta_A^B \delta_C^D$, $\Sigma_{MAB}\Sigma_{MCD} = -2\epsilon_{ABCD}$ and $\Sigma_{MAB}\Sigma_M^{CD} = -4\delta_{AB}^{CD}$. Here $I=(M,X)$ is split as $8\rightarrow 6+2$. Indices $A,B,..$ are $4$ Weyl of $SO(6)$ whereof $\Sigma$ denotes the chirality matrix.} one can show that the Yukawa type coupling can be recast in the ABJM form as well, 
\bea
\L_{Yukawa} &=& -\frac{i}{4} \<X^{\alpha},[X^{\beta},\bar{\psi}_{\dot{\alpha}};\psi_{\dot{\beta}}]\>\Gamma_{I\alpha\dot{\alpha}}\Gamma^{I\dot{\beta}\beta}\cr
&=& \Big(-\frac{i}{2}\epsilon^{ABCD} Z_A^a \bar{\psi}_{Bb} \psi_{Cc} Z_D^d - \frac{i}{2} \epsilon_{ABCD} \bar{\psi}^{Aa} Z_{Bb} Z_{Cc} \psi^{Dd}\cr
&& + i Z_A^a \bar{\psi}_{Bb} Z_c^A \psi^{Bd} - \frac{i}{2} Z_A^a \bar{\psi}_{Bb} Z^B_b \psi^{Ad}\Big)f^{bc}{}_{da}.
\eea
The Chern-Simons term is unchanged as it carries no $SO(8)$ indices. We may also reduce the mass term to corresponding terms in ABJM theory. If we let
\ben
G_{\dot{\beta}}^{\dot{\alpha}} &=& \(\begin{array}{cc}
G^A_B & 0\\
0 & G_A^B
\end{array}\)\label{G2}
\een
we get
\bea
\L_{m} + \L_{flux} &=& -m^2 \<Z^A,Z^A\> + 2 m G^B_A \<Z^A,[Z^B,Z^C;Z^C]\>
\eea
Notice that the second term equals the two terms obtained in \cite{Lee:2009mm} upon expanding the three-bracket in a matrix realization.

\section{Maximally supersymmetric vacuum}\label{Maximally supersymmetric vacuum}
The static Lagrangian can be expressed as a perfect square,
\bea
\L &=& -\<W^{AB}_C,W^{AB}_C\>
\eea
where
\bea
W^{AB}_C &=& \delta^{[A}_C [Z^{B]},Z^C;Z^C] + [Z^A,Z^B;Z^C] + m G^{[A}_C Z^{B]}.
\eea
This shows that the energy is bounded from below by zero, and that it is zero if and only if 
\bea
W^{AB}_C = 0.
\eea
Any space-time independent solution where the gauge field vanishes, and fermions vanish, is a maximally supersymmetric vacuum since the supersymmetry variation is given by $\delta \psi_A = \epsilon_{AB} W^{AB}_C$. One particularly simple solution is obtained by taking $Z^A = T^A$ where 
\bea
T^A &=& \(\begin{array}{c}
T^a\\
T_{\dot{a}}
\end{array}\).
\eea
and $T_{\dot{a}} = 0$, and
\ben
[T^a,T^b;T^c] &=& -m\(\delta^a_c T^b - \delta^b_c T^a\).\label{vacuumeq}
\een
This equation can be solved by taking $T^a \propto \G^a$ which describe $S^3$ with radius 
\ben
R &=& \sqrt{(N-1) m}\label{flux}
\een
We will keep the radius $R$ fixed. The mass parameter $m$ will not play any role for us, but is just a tool we use to obtain the desired geometry characterized by $R$. The radius characterizes the geometry of the M5 brane, and should be kept fixed as $N$ may be taken to infinity. 

The background shall satisfy the vacuum equation (\ref{vacuumeq}) and be valued in the three-algebra
\bea
T^{A} &=& T^{A}_{\vec{b}\vec{b'}} \T^{\vec{b}} \otimes \T^{\vec{b'}}.
\eea
One solution is to take
\bea
T^{A} &=& T^{A}_b \G^b \otimes 1
\eea
where $T^{A}_b \propto \delta^{A}_b$ (for $A = a$) and $\G^b$ is given by (\ref{g}). However, this is an element in the $S^3/{\mb{Z}_K}$-three-algebra only for $K=1$ where the function $1$ can be expressed as a linear combination of $\T^{\vec{a}}$. For generic $K$ we require
\bea
T^{A}(\psi + 2\pi) &=& e^{\frac{2\pi i}{K}} T^{A}(\psi).
\eea
Dimensional reduction amounts to pick only the zero mode\footnote{The name zero mode refers to that we take $m=0$ in Eq (\ref{nakwoo}), an assuption that strictly speaking can be justified only when $K$ is larger than $N$. Since $N$ is taken to infinity, we must take $K$ to infinity as well, such that the ratio $N/K > 1$.}
\bea
\T^{\vec{a}} &=& e^{i\psi} \t \T^{\vec{a}}.
\eea
We will assume the same orbifolding for algebra $\B$, so three-algebra generators in $\A\otimes \B$ will be on the form
\bea
\T^{\vec{a}\vec{a'}} &=& e^{i\psi} \otimes e^{i\psi'} \t \T^{\vec{a}\vec{a'}}
\eea
We want the resulting theory to dimensionally reduce to super Yang-Mills theory when we take large $K$. This means that we can not allow for a three-algebra to survive dimensional reduction. A natural way of reducing three-algebra $\B$ to a Lie algebra is by demanding orbifolding by $\mb{Z}_K$. Assume we took a different discrete group $\mb{Z}_L$ instead, and defined algebra $\B$ on $S^3/{\mb{Z}_L}$. We would then require that
\bea
\frac{2\pi m}{K} + \frac{2\pi n}{L} = \frac{2\pi p}{K}
\eea
As we run through all the integers $m$, $n$, the integer $p$ must also run through all integers. But this is possible only if $L=K$.

We will take the background to be given by
\ben
T^{A} &=& T^{A}_{b} \t \G^b e^{i\psi} \otimes e^{i\psi'} \label{background}
\een
The most general ansatz we can make that satisfies the vacuum equation (\ref{vacuumeq}) is on the form $T^{A}_{bb'}\G^b \otimes \T^{b'}$ where $\T^{b'}$ depends on any two (out of three) coordinates, which makes the three-bracket $[\T^{b'},\T^{c'};\T^{d'}]$ vanishing, which is a necessary condition for the vacuum equation to be satisfied. When we dimensionally reduce we want the background to be commuting so we will also demand that $[\t T^{a'},\t T^{b'}] = 0$. This implies that $\t T^{a'}$ can only depend on one coordinate, and this one coordinate must be $\psi'$ for the orbifolding condition to be satisfied, and we end up with (\ref{background})

However the function $e^{i\psi'}$ does not correspond to a three-algebra element for finite $N$. It corresponds to the unit $N\times N$ matrix, but the three-algebra consists of $N\times (N-1)$ matrices. We thus have an error of order $1/N$ in all our calculations.

\section{D4 from ABJM}\label{D4 from ABJM}
We will now expand the star-three-product ABJM or star-three-product BLG Lagrangian (which formulation we use is just a matter of taste as they are the same) around the supersymmetric vacuum $S^3/{\mb{Z}_K}$ and take the limit $K\rightarrow \infty$ and $R\rightarrow \infty$ while keeping $R/K$ finite. We can also work with the usual matrix realized ABJM theory and expand this theory about the $S^2$ base manifold and get exactly the same result. This is so because for $K>N$, star-three-product ABJM becomes isomorphic to usual ABJM. In the present case $N = N_{\A}N_{\B}$ where $N_{\A}$ is the number of D2 branes and $N_{\B}$ is the number of D4 branes.\footnote{We count the number of D-branes rather than M-branes.}  We will eventually take $N_{\A}\rightarrow \infty$ but it seems plausible that we may take the limit $K\rightarrow \infty$ first, or in other words always secure that $K>N$. For the purpose of deriving D4 we would then only need usual ABJM theory. However to derive the M5, or more generally, to consider cases where $K<N$, we must use star-three-product BLG theory. 

We begin this section by studying the Higgs mechanism in abstract ABJM theory. By abstract, we mean that we keep the realization of the three-algebra unspecified. Working at this abstract level has the advantage that we can apply the same equations to all kind of realizations later on. Next we will expand the resulting abstract Lagrangian in fluctuation fields and derive the full non-Abelian five-dimensional super Yang-Mills Lagrangian and identify the super Yang-Mills coupling constant.

\subsection{The Higgs mechanism}
The vacuum we have found has non-vanishing vacuum expectations values of the scalar fields,
\bea
Z^A &=& T^A + Y^A
\eea
where $T^A$ is the vacuum expectation value, and $Y^A$ are fluctuations. The Higgs mechanism, by which is meant the derivation of an effective action by expanding about a vacuum expecation value, can be studied as a separate problem by itself \cite{Mukhi:2008ux}. The Higgs mechanism does not have to be related with the deconstruction of D4, but arises naturally in deconstruction of D4. 

For non-degenerate situations (meaning square matrices), hermitian three-algebra generators can be diveded into two sets by extracting one generator, let us denote that one as $T^{\sharp}$. Then the remaining generators $T^a$, with $a\neq \sharp$ can be assumed to be hermitian. It follows from the fundamental identity, that the bracket 
\bea
[T^a,T^b] &=& [T^a,T^b;T^{\sharp}]
\eea
is a Lie bracket. For example, for $SO(4)$ we can take $T^a$ to be the Pauli sigma matrices (which are hermitian Lie algebra generators), and $T^{\sharp} = i \mb{I}$. 

If we realize the inner product by matrices, we will use the normalization
\bea
\<T^a,T^b\> &=& \frac{1}{N}\tr(T^a T_b).
\eea
If we use star-product the inner product is 
\bea
\<\T^a,\T^b\> &=& \frac{1}{\pi R^2} \int d^2\sigma \sqrt{G} \T^a * \T_b.
\eea
Here the star-product is superfluous since all higher order terms amount to terms that are total derivatives.

The abstractly defined ABJM Lagrangian then, is given by 
\bea
\L &=& \L_{CS} + \frac{KN}{2\pi} \big\{-\<D_{\mu}Z^A,D^{\mu}Z^A\> - V(Z)\big\}
\eea
\bea
\L_{CS} &=& \frac{KN}{2\pi}\Big\{\frac{1}{2} \epsilon^{\mu\nu\lambda} \Big(\<T^b,[T^c,T^d;T^a]\> A_{\mu}{}^c{}_b \partial_{\nu} A_{\lambda}{}^d{}_a \cr
&& - \frac{2}{3} \<[T^a,T^c;T^d],[T^f,T^b;T^e]\>A_{\mu}{}^b{}_a A_{\nu}{}^d{}_c A_{\lambda}{}^f{}_e\Big)\Big\}
\eea
where
\bea
V(Z) &=& \frac{2}{3}\(\<[Z^A,Z^B;Z^C],[Z^A,Z^B;Z^C]\> - \frac{1}{2}\<[Z^C,Z^A;Z^C],[Z^B,Z^A;Z^B]\>\).
\eea
The gauge covariant derivative is given by
\bea
D_{\mu}Z^A &=& \partial_{\mu}Z^A + [Z^A,T^c;T^d]A_{\mu}{}^d{}_c
\eea
We may expand out the three-bracket and we have
\bea
D_{\mu}Z^A &=& \partial_{\mu}Z^A + Z^A A^R_{\mu} - A^L_{\mu} Z^A
\eea
where
\bea
A^R_{\mu} &=& A_{\mu}{}^d{}_c T_d T^c,\cr
A^L_{\mu} &=& A_{\mu}{}^d{}_c T^c T_d.
\eea
Then the Chern-Simons term can be written\footnote{Here we lend the bracket from the three-algebra and just remove the comma. We write it as $\<\bullet\>$ with no comma. In our two examples this bracket is given by either $\tr$ or $\int$. However in a more abstract setting a comma could be desired. It should be possible to re-introduce a comma in this bracket and hence promote it to a trace form. But this will then be a trace form on the Lie algebra associated with the three-algebra. This is natural since the gauge field takes values in the Lie algebra.}
\bea
\frac{KN}{4\pi}\<A^R dA^R + \frac{2}{3}(A^R)^3 - A^L dA^L - \frac{2}{3}(A^L)^3\>
\eea
In order to study the Higgs mechanism, we define
\bea
a_{\mu} &=& A_{\mu}{}^d{}_c [T^c, T_d],\cr
b_{\mu} &=& A_{\mu}{}^d{}_c \{T^c, T_d\}.
\eea
In terms of these gauge potentials, we have 
\bea
D_{\mu} Z^A &=& \partial_{\mu} Z^A + \frac{1}{2}[Z^A,b_{\mu}] - \frac{1}{2}\{Z^A,a_{\mu}\}
\eea
When we expand about a Higgs vacuum expectation value as
\bea
Z^A &=& T^A + Y^A
\eea
it is natural to define a covariant derivative as
\bea
D_{\mu} Y^A &=& \partial_{\mu}Y^A + \frac{1}{2}[Y^A,b_{\mu}]
\eea
and we have
\bea
D_{\mu} Z^A &=& D_{\mu} Y^A - T^A a_{\mu}
\eea
The Lagrangian reads
\bea
\L &=& \frac{KN}{2\pi}\<-\frac{1}{2} a g + \frac{1}{12}a^3\>\cr
&& + \frac{KN}{2\pi} \Big\{-\<D_{\mu}Y^A,D^{\mu}Y^A\> + \<a_{\mu} D^{\mu} (T^A Y_A + T_A Y^A)\> - T^A T_A \<a_{\mu} a^{\mu}\> - V(T+Z)\Big\}
\eea
where
\bea
g &=& db - \frac{1}{2}b^2
\eea
For later convenience, we make the replacement 
\bea
Z^A &\rightarrow & \frac{1}{\sqrt{-\hbar}}Z^A
\eea
which amounts to $T^A \rightarrow \frac{1}{\sqrt{-\hbar}}T^A$ and $Y^A \rightarrow  \frac{1}{\sqrt{-\hbar}}Y^A$. We recall that $-\hbar>0$. We now get
\bea
\L &=& \frac{KN}{2\pi\hbar}\Big\{-\frac{\hbar}{4}\epsilon^{\mu\nu\lambda}\<a_{\mu} g_{\nu\lambda} + \frac{1}{3}a_{\mu}a_{\nu}a_{\lambda}\>\cr
&& +\<D_{\mu}Y^A,D^{\mu}Y^A\> - \<a_{\mu} D^{\mu} (T^A Y_A + T_A Y^A)\> + T^A T_A \<a_{\mu} a^{\mu}\> + \frac{1}{\hbar^2} V(T+Y)\Big\}
\eea
After the rescaling by $1/\sqrt{-\hbar}$, we declare that
\bea
T^A T_A &=& \frac{R^2}{2}.
\eea
We see that $a_{\mu}$ is auxiliary and can be integrated out. If $\L = a_{\mu}V^{\mu}+\beta a_{\mu} a^{\mu}$, then integrating out $a_{\mu}$ amounts to $\L = -\frac{1}{4\beta} V_{\mu}V^{\mu}$. Here this gives us
\bea
\L &=& \frac{KN\hbar}{32\pi R^2} \<g_{\mu\nu}g^{\mu\nu}\> + \frac{KN}{2\pi\hbar}\(\<D_{\mu}Y^A D^{\mu}Y_A\> - \frac{1}{2R^2}\<\(T^A D_{\mu}Y_A + T_A D_{\mu} Y^A\)^2\>\)
\eea
If we define
\ben
\t Y^A &=& Y^A - \frac{1}{R^2} T^A\(T_B Y^B + T^B Y_B\),\label{proj}
\een
we have the identity
\bea
D_{\mu}\t Y^A D^{\mu} \t Y_A &=& D_{\mu} Y^A D^{\mu} Y_A - \frac{1}{2R^2}\(T^A D_{\mu}Y_A + T_A D_{\mu}Y^A\)^2
\eea
and hence
\bea
\L &=& \frac{KN\hbar}{32\pi R^2} \<g_{\mu\nu}g^{\mu\nu}\> + \frac{KN}{2\pi\hbar}\(\<D_{\mu}\t Y^A D^{\mu}\t Y_A\> + \frac{1}{\hbar^2}V(T+Y)\)
\eea
We rescale $b_{\mu} = 2B_{\mu}$, and use (\ref{norms}) for the inner product, and insert the value for $\hbar$ given by (\ref{hbar}) and we get
\bea
\L &=& -\frac{K}{16\pi^2 R}G_{\mu\nu}G^{\mu\nu} - \frac{KN^2}{\pi^2 R^5}D_{\mu}\t Y^A D^{\mu}\t Y_A
\eea
where
\bea
D_{\mu} \t Y^A &=& \partial_{\mu} \t Y^A + [\t Y^A,B_{\mu}].
\eea
We have dropped the integration over the two-sphere and now view $\L$ as a five-dimensional Lagrangian. If we consider a tensor product $\A\otimes \B$ of three-algebras, then we evaluate only the inner product associated to algebra $\A$, and the inner product associated with algebra $\B$ remains. This inner product is suppressed in the Lagrangian above, but could have been displayed as $\<\bullet\>_{\B}$. If we have a tensor product of three-algebras $\A$ and $\B$ associated with gauge groups $U(N_{\A})\times U(N_{\A})$ and $U(N_{\B})\times U(N_{\B})$, then we have the three-algebra $\A\otimes \B$ associated with gauge group $U(N)\times U(N)$ where $N = N_{\A}N_{\B}$. The remaining inner product of algebra $\B$ will therefore be isomorphic to the unit normalized trace,
\bea
\<\bullet\>_{\B} &=& \tr.
\eea

For later use, we define a gauge field $A_{\mu}$ as
\bea
B_{\mu} &=& -\frac{\lambda}{\epsilon} A_{\mu} 
\eea
The parameter $\epsilon$ is given by (\ref{epsilon}) and explicitly by (\ref{epsilon1}). We now have reached the final form of our Lagrangian, that will be our starting point for performing fluctuation analysis,
\ben
\tr\[\frac{K}{4\pi^2 R}\(\frac{\lambda}{\epsilon}\)^2\(-\frac{1}{4}F_{\mu\nu}F^{\mu\nu}\) - \frac{KN^2}{\pi^2 R^5} \(D_{\mu}\t Y^A D^{\mu} \t Y_A + \frac{1}{\hbar^2} V(T+Z)\)\]\label{infer}
\een
Here
\bea
F_{\mu\nu} &=& \partial_{\mu}A_{\nu}-\partial_{\nu}A_{\mu} + \frac{\lambda}{\epsilon}[A_{\mu},A_{\nu}],\cr
D_{\mu}\t Y^A &=& \partial_{\mu}\t Y^A + \frac{\lambda}{\epsilon}[A_{\mu},\t Y^A].
\eea

\subsection{Fluctuation analysis}
For the purpose of deconstruction of D4, we need to reconsider the above Higgs mechanism. Let us choose the three-algebra basis as
\bea
T^{aa'} &=& T^a \otimes T^{a'}.
\eea
and let us expand 
\bea
D Z^A &\equiv & d Z^A + [Z^A,B] 
\eea
where 
\bea
B &=& A^{dd'}{}_{cc'} \frac{1}{2}\(T^{cc'} T_{dd'} + T_{dd'}T^{cc'}\),\cr
Z^A &=& Z^A_{bb'} T^{bb'}
\eea

We expand
\bea
[T^{bb'},T_{dd'}T^{cc'}] &=& \epsilon\{T^b,T_dT^c\} T^{b'}T_{d'}T^{c'} + T_d T^c T^b [T^{b'},T_{d'}T^{c'}]
\eea
Here
\ben
\epsilon &=& -\frac{i\hbar}{R}.\label{epsilon1}
\een
and the bracket is the Poisson bracket that should not be confused with an anticommutator. We now see that
\bea
D Z^A &=& dZ^A + [Z^A,B] + \epsilon\{Z^A,B\}.
\eea
Expanding about the Higgs vacuum, we keep the following terms,
\ben
D Z^A &=& dY^A + [Y^A,B] + \epsilon\{T^A,B\}\cr
&=& DY^A + \epsilon\{T^A,B\}.\label{deri}
\een
Our previous analysis of the Higgs mechanism must be modified by the addition of the last term, in the deconstruction of D4. 

We may define real coordinates $X^M = (X^A,X^{A+4})$ as
\bea
Z^A &=& \frac{1}{\sqrt{2}} \(X^A + i X^{A+4}\)
\eea
and we have
\bea
Z^A Z_A + Z_A Z^A = R^2 = X^M X^M.
\eea
We split $M = (m,I)$ and we have the metric
\bea
ds^2 = dX^M dX^M = G_{mn} d\sigma^m d\sigma^n + (d\psi + A)^2 + dR^2 + \delta_{IJ} dX^I dX^J
\eea
where the first two terms correspond to the metric on the three-sphere and $G_{mn}$ is the metric on the two-sphere. We now have
\ben
\partial_{(m} Z^A \partial_{n)} Z_A &=& \frac{1}{2} G_{mn},\cr
\partial_{(I} Z^A \partial_{J)} Z_A &=& \frac{1}{2} \delta_{IJ}.\label{metri}
\een

Things get more transparent if we use the notion of fake BLG theory. Then we consider eight-component scalar fields, and fluctuations
\bea
Y^{\alpha} &=& \(\begin{array}{c}
Y^A\\
Y_A
\end{array}\)
\eea
We have a projection, corresponding to (\ref{proj}),
\bea
\t Y^{\alpha} &=& \(\delta^{\alpha}_{\beta} - \frac{1}{R^2} T^{\alpha} T_{\beta}\) Y^{\beta}
\eea
We will decompose the fluctuation part into transverse and tangential parts,
\bea
\t Y^{\alpha} &=& Y^m \partial_m T^{\alpha} + Y^I \partial_I T^{\alpha}.\label{D4fluctuation_expansion}
\eea
Here $\sigma^m$ are coordinates on $S^2$ and $T^I$ are coordinates transverse to $S^2$ and to $R$. It means that one of the directions labeled by $I=1,...,5$ must in fact be along the fiber of $S^3$ as it can not be a radial direction that is projected out by the Higgs mechanism. We relate the fluctuations $Y^m$ and $Y^I$ to a gauge field $A_m$ and five scalar fields $\phi^I$ on D4 according to 
\ben
Y^m &=& \lambda \sqrt{G} \epsilon^{mn} A_n\label{dualize}\\
Y^I &=& \lambda \phi^I.
\een
The constant $\lambda$ is determined by relating Dirac charge quantization of the gauge field, with the winding number of the reparametrization, characterized by large fluctuations $Y^m$. However, for the purpose of determining the super Yang-Mills coupling constant, $\lambda$ does not play any role since we can always make any kind of field redefinition and in particular we can make any field rescaling. This does not change the Yang-Mills coupling constant. So we will not need the actual value of $\lambda$ since it will cancel out in the computation of the Yang-Mills coupling constant. We have put the computation of $\lambda$ in Appendix \ref{appC}, as it may come to use in future studies.

\subsection{The kinetic term}
Let us first consider the kinetic term for the matter fields. We first expand the covariant derivative. From (\ref{deri}) we get
\bea
D_{\mu}Z^{A} &=& \lambda \(\sqrt{G}\epsilon^{mn} F_{\mu n} \partial_m T^A + D_{\mu}\phi^I \partial_I T^A\). 
\eea
where
\bea
D_{\mu}\phi^I &=& \partial_{\mu}\phi^I + \frac{\lambda}{\epsilon} [A_{\mu},\phi^I],\cr
F_{\mu n} &=& \partial_{\mu}A_n - \partial_n A_{\mu} + \frac{\lambda}{\epsilon} [A_{\mu},A_n].
\eea
The kinetic term becomes
\ben
-\<D_{\mu}Z^A,D^{\mu}Z^A\> &=& \lambda^2\(-\frac{1}{2} \<F_{\mu n},F^{\mu n}\> - \frac{1}{2} \<D_{\mu}\phi^I,D^{\mu}\phi^I\>\).\label{kin}
\een
Here we have used the metric components (\ref{metri}).

\subsection{The sextic potential plus flux term}

\subsubsection{Quadratic order}
At quadratic order we need to combine the contribution from the sextic potential and the flux term to get
\bea
\lambda^2 R^2 |\epsilon|^2 \mu \(-\frac{1}{4}\<f_{mn},f^{mn}\> - \frac{1}{2}\<\partial_m \phi^I,\partial^m\phi^I\>\)
\eea
where 
\bea
f_{mn} &=& \partial_m A_n - \partial_n A_m.
\eea

\subsubsection{Cubic order}
At cubic order and higher, the flux term drops to zero as we take $N$ large, as can be inferred from Eq (\ref{flux}). We thus only need to expand the sextic potential. We choose to work with the BLG theory sextic potential which is more convenient than the ABJM sextic potential. We first expand 
\bea
\<[\T^{\alpha},\T^{\beta};Y^{\gamma}],[\T^{\alpha},Y^{\beta};Y^{\gamma}]\>
\eea
and then we note that there are $12$ terms that give the same contribution so it will be sufficient to just compute this term and multiply the result by $12$. We assume that 
\bea
\partial_{\psi} Y^{\gamma} &=& 0,\cr
T^{\alpha} T_{\alpha} &=& R^2,
\eea
where $\psi$ parametrizes the fiber on $S^3/{\mb{Z}_K}$. We expand 
\bea
[\T^{\alpha},\T^{\beta};Y^{\gamma}] &=& [\T^{\alpha},Y_{\gamma}]\T^{\beta}-[\T^{\beta},Y_{\gamma}]\T^{\alpha}+Y_{\gamma}[\T^{\alpha},\T^{\beta}],\cr
[\T^{\alpha},Y^{\beta};Y^{\gamma}] &=& [\T^{\alpha},Y_{\gamma}]Y^{\beta}-[Y^{\beta},Y_{\gamma}]\T^{\alpha}+Y_{\gamma}[\T^{\alpha},\T^{\beta}]
\eea
If $Y_{\gamma}$ is star-three-multiplied with a commutator reduced to the $S^2$ base-manifold, we will necessarily need to act by a $\psi$-derivative on $Y_{\gamma}$ which kills the whole term. As for the second expansion, we are interested in only the non-Abelian part as the Abelian part will vanish in the large $R$ limit. In the above expansions we thus keep the terms
\bea
[\T^{\alpha},\T^{\beta};Y^{\gamma}] &=& [\T^{\alpha},Y_{\gamma}]\T^{\beta}-[\T^{\beta},Y_{\gamma}]\T^{\alpha},\cr
[\T^{\alpha},Y^{\beta};Y^{\gamma}] &=& -[Y^{\beta},Y_{\gamma}]\T^{\alpha}.
\eea
and we get
\bea
\<[\T^{\alpha},\T^{\beta};Y^{\gamma}],[\T^{\alpha},Y^{\beta};Y^{\gamma}]\> &=& R^2\<[\T^{\beta},Y_{\gamma}],[Y^{\beta},Y_{\gamma}]\>\cr
&=& R^2 \epsilon\sqrt{G}\epsilon^{mn} \<\partial_m \T^{\beta} \partial_n Y_{\gamma},[Y^{\beta},Y_{\gamma}]\>.
\eea
We next insert the expansion Eq (\ref{D4fluctuation_expansion}) and we get
\bea
&=& R^2 \epsilon\sqrt{G}\epsilon^{mn} G_{mp}\<\partial_n Y_{\gamma},[Y^p,Y_{\gamma}]\>\cr
&=& R^2 \lambda \epsilon G^{mn} \<\partial_m Y_{\gamma},[A_n,Y_{\gamma}]\>\cr
&=& \lambda^3 R^2\epsilon\Bigg(G^{mn}G^{pq} \<\partial_m A_p,[A_n,A_q]\> + G^{mn}\<\partial_m \phi^I,[A_n,\phi^I]\>\Bigg)
\eea
In the last step we have noted a cancelation of two terms
\bea
\sqrt{G}\epsilon^{mn} \<A_n,[A_m,Y^R]\> -  \sqrt{G}\epsilon^{mn}\<Y^R,[A_n,A_m]\>
\eea
which cancel by trace invariance $\<X,[Y,Z]\> = \<[X,Y],Z\>$ and cyclicity $\<X,Y\> = \<Y,X\>$.

\subsubsection{Quartic order}
At quartic order we have three terms of the type
\bea
&&\<[\T^{\alpha},Y^{\beta};Y^{\gamma}],[\T^{\alpha},Y^{\beta};Y^{\gamma}]\> \cr
&=& R^2\<[Y^{\beta},Y_{\gamma}],[Y^{\beta},Y_{\gamma}]\>\cr
&=& \lambda^4 R^2\Bigg(G^{mp} G^{nq} \<[A_m,A_n],[A_p,A_q]\> + 2 G^{mn} \<[A_m,\phi^I],[A_n,\phi^I]\>\cr
&& + \<[\phi^I,\phi^J],[\phi^I,\phi^J]\>\Bigg)
\eea

\subsection{Summarizing}
We shall multiply the cubic term by $-\frac{12}{12\hbar^2}.\frac{KN^2}{\pi^2 R^5}$ and the quartic term by $-\frac{3}{12\hbar^2}.\frac{KN^2}{\pi^2 R^5}$ as we infer from (\ref{infer}), and the combinatorics give us the factors of $12$ and $3$ respectively. Discarding the common factor of $\frac{KN^2}{\pi^2 R^5}$, we have
\bea
-\frac{R^2 \lambda^2}{2\hbar^2} G^{mn}\Big(|\epsilon|^2 \<\partial_m \phi^I,\partial_n \phi^I\> + \lambda (\epsilon - \epsilon^*) \<\partial_m \phi^I,[A_n,\phi^I]\> + \lambda^2 \<[A_m,\phi^I],[A_n,\phi^I]\>\Big)
\eea
plus
\bea
&&-\frac{R^2\lambda^2}{4\hbar^2} G^{mp}G^{nq} \Big(|\epsilon|^2 \<f_{mn},f^{pq}\> + \frac{1}{2}\lambda (\epsilon - \epsilon^*) \<\partial_m A_n,[A_p,A_q]\> + \lambda^2 \<[A_m,A_n],[A_p,A_q]\>\Big)
\eea
The various terms combine into covariant expressions. Re-instating the common factor of $\frac{KN^2}{\pi^2 R^5}$, we get
\bea
\frac{KN^2}{\pi^2 R^5}\lambda^2 \(-\frac{1}{4} G^{mp}G^{nq} \<F_{mn},F_{pq}\> - \frac{1}{2} G^{mn} \<D_m Y^I,D_n Y^I\>\)
\eea
where 
\bea
D_m Y^I &=& \partial_m Y^I + \frac{\lambda}{\epsilon} [A_m,Y^I],\cr
F_{mn} &=& \partial_m A_n - \partial_n A_m + \frac{\lambda}{\epsilon} [A_m,A_n].
\eea
To read off the Yang-Mills coupling constant, we may make a field redefinition that removes the factor $\frac{\lambda}{\epsilon}$ from the covariant derivative and the gauge field strength, and puts $\(\frac{|\epsilon|}{\lambda}\)^2$ as an overall factor. Then we get the total overall factor as
\bea
\frac{KN^2}{\pi^2 R^5} \lambda^2 \(\frac{|\epsilon|}{\lambda}\)^2 &=& \frac{K}{4\pi^2 R}
\eea
from which we read off the Yang-Mills coupling as 
\bea
g_{YM}^2 &=& 4 \pi^2 \frac{R}{K}.
\eea
We notice that the first term in (\ref{infer}) also comes with the overall factor $\frac{K}{4\pi^2 R}$ after we have multiplied it by $\(\frac{|\epsilon|}{\lambda}\)^2$. Likewise we find this factor in the kinetic term (\ref{kin}). We have now derived the full non-Abelian super Yang-Mills Lagrangian from ABJM theory.

The compactification radius is $\frac{R}{K}$ if we put M five-brane on the orbifold $S^3/{\mb{Z}_K}$. We thus see that we can derive the selfdual coupling constant of M5 brane directly from the M2 brane. We think this is a quite remarkable discovery. It gives us hope that it might be possible to derive M5 brane physics from ABJM theory. 

There are higher order non-Abelian terms induced from ABJM theory, that we did not bother to compute. These are $1/R$-suppressed, and may be neglected for small $g_{YM}$. We also did not bother to compute the $1/R$ correction terms that arise because we consider SYM on $\mb{R}^{1,2} \times S^2$ rather than on $\mb{R}^{1,4}$. In the Abelian case, these terms were considered in \cite{Nastase:2010uy}. Here we considered the flat space limit with both $K,R\rightarrow \infty$ while $g_{YM}$ is kept fixed. 

When we derived the five-dimensional super Yang-Mills Lagrangian, we naturally ended up with star-commutators rather than matrix commutators. But these are isomorpic. For instance in the relation
\bea
[\T^{a'},\T^{b'};\T^{c'}] &=& \T_{c'}[\T^{a'},\T^{b'}] + [\T^{a'},\T_{c'}]\T^{b'} - [\T^{b'},\T_{c'}]T^{a'}
\eea
If we use star-three-product, then upon dimensional reduction these star-commutators are given by
\bea
[T^a,T^b] &=& \epsilon_{\B} \{T^a,T^b\} + \O(\epsilon_{\B}^2)
\eea
But upon dimensional reduction we may also map functions into matrices. In that case we have
\bea
[T^{a'},T^{b'};T^{c'}] &=& T_{c'}[T^{a'},T^{b'}] + [T^{a'},T_{c'}]T^{b'} - [T^{b'},T_{c'}]T^{a'}
\eea
and now the commutators on the right-hand side are usual matrix commutators, so we have the isomporphism
\bea
[T^{a'},T^{b'}] &\cong & [\T^{a'},\T^{b'}]
\eea

\section{Lagrangian for a selfdual three-form}\label{Lagrangian for a selfdual three-form}
We can not write down the action of a selfdual three-form in six dimensions. From the M2 we rather get \cite{Ho:2008ve} (at quadratic order there is no essential difference between Abelian and non-Abelian M5)
\ben
\L &=& -\frac{1}{4}H_{\mu\alpha\beta}H^{\mu\alpha\beta}-\frac{1}{12}H_{\alpha\beta\gamma}H^{\alpha\beta\gamma}-\frac{1}{2}\sqrt{g}\epsilon^{\alpha\beta\gamma}\epsilon^{\mu\nu\lambda}\partial_{\beta}B_{\mu\gamma}\partial_{\nu}B_{\lambda\alpha}.\label{M5}
\een
where $\mu=0,1,2$ and $\alpha=3,4,5$. That is, we break $SO(1,5)\rightarrow SO(1,2)\times SO(3)$. To better understand how to interpret this Lagrangian for the selfdual gauge field we compute its Hamiltonian and compare with the Hamiltonian of a non-chiral gauge field. Let us start with the Lagrangian of a non-chiral gauge field on $\mb{R}\times M_5$,
\bea
\L &=& -\frac{1}{12}H_{MNP}H^{MNP}
\eea
Let us split the vector index as $M = (0,m)$ and compute the conjugate momenta
\bea
E^{mn} = \frac{\partial \L}{\partial \partial_0 B_{mn}} = - \frac{1}{2}H^{0mn}
\eea
and the Hamiltonian
\bea
\H &=& \frac{1}{4}H^{0mn}H^0{}_{mn} + \frac{1}{12} H^{mnp} H_{mnp}
\eea
If we split the field strength into selfdual parts,
\bea
*H^{\pm} &=& \pm H^{\pm}
\eea
we get
\ben
\H &=& \frac{1}{12}\epsilon^{0mnpqr}\(-H_{0mn}^+ H_{pqr}^+ + H_{0mn}^- H_{pqr}^-\).\label{H}
\een

Let us now repeat the same steps, but with the Lagrangian (\ref{M5}). We split $\mu = (0,i)$ and compute the conjugate momenta
\bea
E^{ij} &=& 0,\cr
E^{i\alpha} &=& -\frac{1}{2}\epsilon^{\alpha\beta\gamma}\epsilon^{j0i}\partial_{\beta}B_{j\gamma},\cr
E^{\alpha\beta} &=& -\frac{1}{2}H^{0\alpha\beta}
\eea
and the Hamiltonian
\bea
\H &=& -\frac{1}{4}H^{0\alpha\beta}H_{0\alpha\beta} + \frac{1}{4}H^{i\alpha\beta}H_{i\alpha\beta}+\frac{1}{12}H^{\alpha\beta\gamma}H_{\alpha\beta\gamma}
\eea
Let us assume this field strength is already selfdual. Then we can also write this Hamilonian as
\bea
\H &=& \epsilon^{\alpha\beta\gamma ij0} \(\frac{1}{4}H^+_{0\alpha\beta}H^+_{\gamma ij}-\frac{1}{2}H^+_{i\alpha\beta}H^+_{\gamma j 0} + \frac{1}{12} H^+_{\alpha\beta\gamma}H^+_{0ij}\)\cr
&=& \frac{1}{12}\epsilon^{mnpqr0}H^+_{omn}H^+_{pqr}
\eea
which agrees with the chiral part of the non-chiral Hamiltonian (\ref{H}).

\section{Multiple M5}\label{Multiple M5 in decompactification limit}
Let us choose $K=1$ for simplicity, and consider expanding mass deformed star-three-product BLG theory about $S^3$. We have not settled the issue as to whether we may relax the form of Eq (\ref{red}) and admit a complete set of functions on $S^3$. We can rigorously just work with ABJM theory or any equivalent formulation thereof. We thus can not in any way rigorously deconstruct a six-dimensional theory from BLG/ABJM theory at this stage. Let us suppose that we have overcome this obstacle by means of monopole operators say, and have understood how to do this deconstruction. For example by using our conjectural truncated inner product (\ref{inner}). We may then ask what can be the most general interaction terms in the M5 Lagrangian? When we expand BLG theory, we get terms like
\bea
\t K\((DX)^2 + X^6\) &=& \t K \((DY)^2 + T^4 Y^2 + T^3 Y^3 + T^2 Y^4 + T Y^5 + Y^6\)
\eea
where $X$ and $T$ have length dimensions $L^{-\frac{1}{2}}$, and $\t K$ is some unknown overall coupling coefficient of star-three-product theory. Now let us rescale the field $Y$ into a field $\phi$ with length dimension $L^{-2}$. The field $\phi$ represents one of the scalar fields or a two-form gauge potential in the M5 brane world volume. The M5 brane Lagrangian is now
\bea
(D\phi)^2 + c_3 \phi^3 + c_4 \phi^4 + c_5 \phi^5 + c_6 \phi^6
\eea
where $[c_3] = L^0$, $[c_4] = L^2$, $[c_5] = L^4$, $[c_6] = L^6$. The only length parameter available comes from the radius $R$. In the decompactification limit $R\rightarrow \infty$ causing the interaction terms to blow up to infinity. The only interaction term that could stay finite is on the form $c_3\phi^3$. 
We may drop all interactions and just keep gauge interactions. Indeed this is interesting starting point since it can teach us about the gauge structure of multiple M5. We assume the three-algebra generators are real. We expand the eight real scalar fields $X^I$ about the $S^3$ background, that we write as $T^I$, and define fluctuation fields according to
\bea
X^I &=& T^I + Y^I,\cr
Y^I &=& Y^{\aalpha} \partial_{\aalpha} T^I + Y^J\partial_J T^I,\cr
Y^{\aalpha} &=& \frac{1}{2}\sqrt{g}\epsilon^{\aalpha\bbeta\ggamma}B_{\bbeta\ggamma},\cr
B_{\aalpha\mu} &=& T^c \partial_{\aalpha} T_d A_{\mu}{}^d{}_c,\cr
A_{\mu} &=& T^c T_d A_{\mu}{}^d{}_c
\eea
and
\bea
\Lambda_{\ggamma} &=& -T^c \partial_{\ggamma} T_d \Lambda^d{}_c,\cr
\Lambda &=& T^c T_d \Lambda^d{}_c.
\eea
The easiest thing to deconstruct are the gauge variations. From BLG gauge variation (\ref{deltaX}) we get
\bea
\delta B_{\aalpha\bbeta} &=& \partial_{\aalpha}\Lambda_{\bbeta} - \partial_{\bbeta} \Lambda_{\aalpha} + [B_{\aalpha\bbeta},\Lambda]\cr
\delta Y^I &=& [Y^I,\Lambda].
\eea
and from (\ref{deltaA}) we get
\bea
\delta B_{\aalpha\mu} &=& \partial_{\aalpha}\Lambda_{\mu} - D_{\mu}\Lambda_{\aalpha} + [B_{\aalpha\mu},\Lambda],\cr
\delta A_{\mu} &=& \partial_{\mu}\Lambda + [A_{\mu},\Lambda].
\eea
We should also be able to formulate the gauge variations in an $SO(1,5)$ covariant way. We define the gauge covariant derivative as
\bea
D_M &=& \partial_M + [\bullet,A_M].
\eea
It is now not hard to see that the following variations. 
\ben
\delta B_{MN} &=& D_M \Lambda_N - D_N \Lambda_M + [\Lambda,B_{MN}],\cr
\delta A_M &=& \partial_M \Lambda + [\Lambda,A_M]\label{gaugevariation}
\een
constitute the $SO(1,5)$ covariant counterpart of the above gauge variations. These gauge variations are consistent with assigning $Y^I$, $B_{MN}$, $\Lambda_M$ to be three-algebra valued, and $A_M$, $\Lambda$ to be Lie algebra valued where the Lie algebra is the one that is associated to the three-algebra. Perhaps a bit surprising that the two-form shall be three-algebra valued as this is a gauge field. With this assignment we can also show that these gauge variations close according to
\bea
[\delta_{\Lambda'},\delta_{\Lambda}] &=& \delta_{\Lambda''}
\eea
with 
\bea
\Lambda'' &=& [\Lambda,\Lambda'],\cr
\Lambda''_M &=& [\Lambda,\Lambda'_M] + [\Lambda_M,\Lambda'].
\eea
To show this one may use the generalized Jacobi identities \cite{G} of a three-algebra. One may also assume all the fields are Lie algebra valued. But this seems to be physically incorrect, or at least does not seem to give a theory that can describe the M5 branes. Anyway, gauge symmetry and supersymmetry works out well so at this stage we can not explain why we should not assume all fields be Lie algebra valued. Such a theory can not be obtained from the M2 brane by fluctuation analysis though. 

Closure of these gauge varitions is highly non-trivial. We may in particular notice that the variation of $B_{MN}$ contains a term $[\Lambda,B_{MN}]$ which does not look gauge covariant. The familiar situation is that a variation of a connection one-form is a gauge covariant quantity (and this is indeed also the case here), but with a non-Abelian two-form something much more subtle is apparently going on. 

If we just have gauge interactions, we may easily write down supersymmetry variations
\bea
\delta Y^I &=& i\bar{\omega}\Gamma^I\chi,\cr
\delta B_{MN} &=& i\bar{\omega}\Gamma_{MN}\chi,\cr
\delta A_M &=& 0,\cr
\delta \chi &=& \frac{1}{12}\Gamma^{MNP}\omega H_{MNP}+\Gamma^M\Gamma_I\omega D_M\phi^I
\eea
where
\bea
H_{MNP} &=& D_M B_{NP} + D_P B_{MN} + D_N B_{PM}.
\eea
As we demonstrate in the Appendix \ref{appA}, these supersymmetry variations close up to a gauge variation of precisely the proposed form (\ref{gaugevariation}), on the equations of motion
\bea
F_{MN} &=& 0,\cr
H_{QMN} + \frac{1}{6}\epsilon_{QMNRST}H^{RST} &=& 0,\cr
\Gamma^M D_M \chi &=& 0.
\eea

In \cite{Lambert:2010wm} an additional field $C^{M}$ was introduced in order to admit more general interaction terms and a clear relation with the theory of D4. No field such as $C^{M}$ arises in the fluctuation analysis of M2. However $A_{M}$ and $B_{MN}$ both arise naturally as fluctuation fields. By solving for the equation of motion for $C^{M}$ in \cite{Lambert:2010wm}, it was found that the theory becomes equivalent with D4. Our proposal for the theory of multiple M5 is somewhat related with \cite{Lambert:2010wm}. If we put $C^{M} = 0$ in \cite{Lambert:2010wm} we arrive at the above supersymmetry variations. 

In \cite{Lambert:2010wm} it was claimed that the theory with $C^{M} = 0$ corresponds to non-interacting M5 branes. We may move the branes transverse to each other with no energy cost since there is no scalar potential. That alone does not mean the branes are non-interacting. If the branes are non-interacting then also small fluctuations on one brane should not affect the other brane. This may not be true if there are non-Abelian gauge interactions. If we separate two M5 branes by giving an expectation value to a scalar field, then we induce a Higgs mass to the gauge field and to all the other fluctuation fields. If we are able to find a selfdual string soliton, its tension should be determined by the separation of the M5 branes. By scattering elementary fluctuation quanta againts the string, it should produce a wave that goes out to the other M5 brane, and the two M5 branes would be interacting.  

While it appears that having only gauge interactions give consistent supersymmetry variations that close on-shell, it also appears that these equations of motion can not follow from an action. To this end it seems one needs to introduce a selfdual auxiliary three-form in order to be able to write down a Chern-Simons type of action for the connection one-form $A_M$. This is work in progress.

\vskip 2truecm

\subsubsection*{Acknowledgements}I have discussed this work with Takao Suyama and Soo-Jong Rey. This work was supported by the National Research Foundation of Korea(NRF) grant funded by the Korea government(MEST) through the Center for Quantum Spacetime(CQUeST) of Sogang University with grant number 2005-0049409.

\newpage

\appendix
\section{Closing M5 supersymmetry}\label{appA}
Here we demonstrate closure of the supersymmetry variations 
\bea
\delta Y^I &=& i\bar{\omega}\Gamma^I\chi,\cr
\delta B_{MN} &=& i\bar{\omega}\Gamma_{MN}\chi,\cr
\delta A_M &=& 0,\cr
\delta \chi &=& \frac{1}{12}\Gamma^{MNP}\omega H_{MNP}+\Gamma^M\Gamma_I\omega D_M\phi^I
\eea
We define
\bea
\Gamma &=& \Gamma_{012345},\cr
\epsilon_{012345} &=& 1
\eea
and assume that 
\bea
\Gamma \omega &=& -\omega,\cr
\Gamma \chi &=& \chi.
\eea
Using the gamma matrix identities
\bea
\{\Gamma_{MN},\Gamma^{PQR}\} &=& -12 \delta_{MN}^{[PQ}\Gamma^{R]} + 2 \epsilon_{SMN}{}^{PQR} \Gamma^S \Gamma,\cr
\Gamma^{MNP}\Gamma_Q\Gamma_{MN} &=& -24 \delta_Q^P + 4 \Gamma_Q \Gamma^P
\eea
and for $\Gamma \omega = -\omega$ and $\Gamma \rho = -\rho$ we have the Fierz identity
\bea
\omega\bar{\rho} - \rho\bar{\omega} &=& \(-\frac{1}{16}(\bar{\rho}\Gamma_{M}\omega)\Gamma^{M}+\frac{1}{16}(\bar{\rho}\Gamma_{M}\Gamma_A\omega)\Gamma^{M}\Gamma^A\)(1+\Gamma)\cr
&&-\frac{1}{192}(\bar{\rho}\Gamma_{MNP}\Gamma_{AB}\omega)\Gamma^{MNP}\Gamma^{AB}
\eea
we get
\bea
[\delta_{\rho},\delta_{\omega}]Y^I &=& -2 i\bar{\omega} \Gamma^Q \rho D_Q Y^I,\cr
[\delta_{\rho},\delta_{\omega}]B_{MN} &=& -2 i\bar{\omega} \Gamma^Q \rho D_Q B_{MN},\cr
[\delta_{\rho},\delta_{\omega}]A_M &=& 0.
\eea
Using gamma matrix relations
\bea
\Gamma_{MN}\Gamma^{PQR} + \Gamma^{PQR}\Gamma_{MN} &=& -12 \delta_{MN}^{[PQ} \Gamma^{R]} + 2 \Gamma_{MN}{}^{PQR},\cr
\Gamma_{MNPQR} &=& \epsilon_{MNPQRS}\Gamma\Gamma^S
\eea
and
\bea
\Gamma^{MNP} \Gamma^Q \Gamma_{MN} &=& -24\eta^{PQ} + 4 \Gamma^Q \Gamma^P,\cr
\Gamma^{MNP}\Gamma^{QRS}\Gamma_{MN} &=& 4\(3 \eta^{P[Q,RS]} + 3 \eta^{P[Q} \Gamma^{RS]} + \Gamma^{PQRS}\),\cr
\Gamma^{PQRS} + \Gamma^{QRS}\Gamma^P &=& 3 \eta^{P[Q}\Gamma^{RS]},\cr
\Gamma^M \Gamma^{QRS} + \Gamma^{QRS} \Gamma^M &=& 6 \eta^{M[Q} \Gamma^{RS]}
\eea
we find explicitly
\bea
[\delta_{\rho},\delta_{\omega}] B_{MN} &=& -2i \bar{\omega}\Gamma^T \rho \partial_T B_{MN}\cr
&& + 2D_{[M} \Lambda_{N]} + [\Lambda,B_{MN}]\cr
&&+ i\bar{\omega} \Gamma^T \rho \(H_{TMN} - \frac{1}{6}\epsilon_{TMN}{}^{PQR} H_{PQR}\),\cr
[\delta_{\rho},\delta_{\omega}] A_M &=& -2i\bar{\omega}\Gamma^T \rho \partial_T A_M \cr
&&+ D_M\Lambda \cr
&&+ 2i \bar{\omega}\Gamma^T \rho F_{TM},\cr
[\delta_{\rho},\delta_{\omega}] \phi^I &=& -2i \bar{\omega}\Gamma^T \rho \partial_T \phi^I\cr
&&+ [\Lambda,\phi^I],\cr
[\delta_{\rho},\delta_{\omega}] \chi &=&  -2i \bar{\omega}\Gamma^T \rho \partial_T \chi\cr
&&+ [\Lambda,\chi]\cr
&&+ \frac{3i}{4} (\bar{\omega}\Gamma^N \rho) \Gamma_N \(\Gamma^M D_M \chi\)\cr
&&+ \frac{i}{2} (\bar{\rho}\Gamma^N \Gamma^A \omega) \Gamma_N \Gamma_A \(\Gamma^M D_M \chi\)
\eea
where 
\bea
\Lambda_N &=& -i\bar{\omega} \Gamma^T \rho B_{NT} - 2i \bar{\omega} \Gamma_N \Gamma_I \rho \phi^I,\cr
\Lambda &=& 2i \bar{\omega} \Gamma^T \rho A_T.
\eea
Interestingly we have closure up to a gauge variation of precisely the form (\ref{gaugevariation}), provided we go on-shell where we have the equations of motion
\bea
F_{MN} &=& 0,\cr
\Gamma^M D_M \chi &=& 0\cr
H_{QMN} + \frac{1}{6}\epsilon_{QMNRST}H^{RST} &=& 0.
\eea

\newpage 

\section{Quantization condition for the fluctuation fields}\label{appC}
Since the gauge field is associated with a Dirac quantization condition, which in the Abelian case reads
\bea
\int \frac{F}{2\pi} &\in & \mb{Z}
\eea
we should find that its dual field $Y^m$ is also subject to a quantization condition. 

Let us assume that the relation
\bea
Y^{\alpha} &=& Y^m \partial_m T^{\alpha}
\eea
holds for any finite $Y^m$. Let us map our complex coordinates into three real coordinates, so that
\bea
T^i T^i &=& R^2
\eea
describes $S^2$ embedded in flat euclidean three-dimensional space. 

We are now particularly interested in coordinate transformations that are not continuosly connected with the identity. On a circle such transformations are characterized by a winding number, which can be any integer number. The winding number of the coordinate transformation 
\bea
\varphi \mapsto \varphi' = \varphi + Y^{\varphi}
\eea
is computed intrinsically by the integral
\bea
w = \int \frac{d\varphi'}{2\pi} = 1 + \int \frac{d Y^{\varphi}}{2\pi}. 
\eea
We can express the same thing extrinsically by the integral
\bea
w &=& \int \frac{1}{2\pi R^2} \epsilon_{ij} X^i dX^j
\eea
where 
\bea
X^i = T^i + Y^i
\eea
is a coordinate transformation that respects the constraint $T^i T^i = R^2 = X^i X^i$. We get back the intrinsic integral if we let
\bea
T^1 + i T^2 &=& R e^{i\varphi},\cr
X^1 + i X^2 &=& R e^{i\varphi'}.
\eea
There is a third way of expressing the winding number. Let us define
\bea
Y^i &=& Y^{\varphi} \partial_{\varphi} T^i
\eea
If we then expand out the extrinsic integral we can compute the variation of the winding number (that is, $\delta w = w-1$ where $w=1$ in the original configuration) as
\bea
\delta w &=& \int \frac{1}{2\pi R^2} \epsilon_{ij} \(Y^i dT^j + T^i dY^j + Y^i dY^j\)\cr
&=& \int \frac{d\varphi}{2\pi R^2} \epsilon_{ij} T^i dT^j \partial_{\varphi} Y^{\varphi}
\eea
To obtain this result we have used
\bea
T^i \partial_{\varphi} T^i &=& 0,\cr
\partial_{\varphi}^2 T^i &=& -T^i
\eea
as is explicit from the parametrization above.

\subsection{Two-sphere}
This can be generalized to $S^n$ for any integer $n$. For $S^2$ we compute the winding number by the extrinsic integral
\bea
w &=& \int \frac{1}{8\pi R^3} \epsilon_{ijk} X^i dX^j \wedge dX^k.
\eea
Defining the variation
\bea
X^i &=& T^i + Y^i,\cr
Y^i &=& Y^m \partial_m T^i
\eea
we get, by doing the same steps as we did above for $S^1$,
\bea
\delta w &=& \int \frac{1}{8\pi R^3} \epsilon_{ijk} T^i dT^j \wedge dT^k D_m Y^m.
\eea
To get here we have used 
\bea
D_m \partial_n T^i = -\frac{1}{R^2} G_{mn} T^i.
\eea
We can also express this integral as
\bea
\delta w &=& \frac{1}{4\pi}\int d\varphi \wedge d\theta \sin \theta  D_m Y^m
\eea
Let us denote the metric of the unit sphere as $\hat{G}_{mn}$. Then we have
\bea
\delta w &=& \int d^2 \sigma \sqrt{\hat{G}} D_m Y^m.
\eea
We dualize $Y^m$ according to Eq (\ref{dualize}). We then get
\bea
\delta w &=& \frac{\lambda}{4\pi R^2} \int F.
\eea
Both sides are integer quantized, and this fixes
\ben
\lambda &=& 2R^2.\label{twosphere}
\een

\subsection{Three-sphere}
Let us do one more example, that perhaps can be for future use to derive the M5 brane coupling directly from M2 without taking the detour via five-dimensional super Yang-Mills. Let us consider $S^3$ and dualize the fluctuations as
\ben
Y^{\aalpha} &=& \lambda \sqrt{g}\epsilon^{\aalpha\bbeta\ggamma} B_{\bbeta\ggamma}.\label{parent}
\een
The integer winding number is given by
\bea
w &=& \int \frac{1}{12\pi^2 R^4} \epsilon_{ijkl} T^i dT^j\wedge dT^k \wedge dT^l
\eea
and for the difference that we define via the fluctuation field as
\bea
Y^i &=& T^i + Y^{\aalpha} \partial_{\aalpha} T^i
\eea
we get
\bea
\delta w &=& \int \frac{1}{12\pi^2 R^4} \epsilon_{ijkl} T^i dT^j\wedge dT^k \wedge dT^l D_{\aalpha} Y^{\aalpha}
\eea
or in terms of local coordinates,
\bea
\delta w &=& \frac{1}{2\pi^2 R^3} \int d^3 \sigma \sqrt{g} D_{\aalpha} Y^{\aalpha}
\eea
Inserting the ansatz, we get
\bea
\delta w &=& \frac{\lambda}{2\pi^2 R^3} 2\int H
\eea
and we see that we must choose
\ben
\lambda &=& \frac{\pi R^3}{2}\label{threesphere}
\een
to be compatible with the Dirac charge quantization condition $\int H \in 2\pi \mb{Z}$.

\subsection{Three-sphere as a fiber bundle over two-sphere}
Let us establish that (\ref{threesphere}) and (\ref{twosphere}) are related by
\bea
\int d\psi B_{m\psi} &=& A_m
\eea
in dimensional reduction where we put $\partial_{\psi} = 0$ and we thus have 
\bea
2\pi B_{m\psi} &=& A_m.
\eea
Inserting this into (\ref{parent}) we get
\bea
Y^m &=& 2\frac{\pi R^3}{2}\sqrt{g}\epsilon^{mn\psi} B_{n\psi}\cr
&=& \frac{R^2}{2}\sqrt{G}\epsilon^{mn}A_n\cr
&=& 2\(\frac{R}{2}\)^2 \sqrt{G}\epsilon^{mn}A_n.
\eea
Then we must recall that the radius on the base $S^2$ is $\frac{R}{2}$. We see that we reproduce the result for the two-sphere in Eq (\ref{dualize}), (\ref{twosphere}).


\newpage

\begin{thebibliography}{999}

\bibitem{Tong:2005un}
  D.~Tong,
  arXiv:hep-th/0509216.

\bibitem{ABJM}
  O.~Aharony, O.~Bergman, D.~L.~Jafferis and J.~Maldacena,
  arXiv:0806.1218 [hep-th].


\bibitem{Kim:2009wb}
  S.~Kim,
  Nucl.\ Phys.\  B {\bf 821}, 241 (2009)
  [arXiv:0903.4172 [hep-th]].

\bibitem{Bashkirov:2010kz}
  D.~Bashkirov and A.~Kapustin,
  arXiv:1007.4861 [hep-th].

\bibitem{Kapustin:2010xq}
  A.~Kapustin, B.~Willett and I.~Yaakov,
  JHEP {\bf 1010}, 013 (2010)
  [arXiv:1003.5694 [hep-th]].

\bibitem{Drukker:2010nc}
  N.~Drukker, M.~Marino and P.~Putrov,
  arXiv:1007.3837 [hep-th].




\bibitem{Schnabl:2008wj}
  M.~Schnabl and Y.~Tachikawa,
  JHEP {\bf 1009}, 103 (2010)
  [arXiv:0807.1102 [hep-th]].


\bibitem{G}
  A.~Gustavsson,
  arXiv:0709.1260 [hep-th].



\bibitem{BLG}
  J.~Bagger and N.~Lambert,
  Phys.\ Rev.\  D {\bf 77}, 065008 (2008)
  [arXiv:0711.0955 [hep-th]],

\bibitem{BL}
  J.~Bagger and N.~Lambert,
  arXiv:0807.0163 [hep-th].

\bibitem{Gustavsson:2010yr}
  A.~Gustavsson,
  arXiv:1012.4568 [hep-th].



\bibitem{Lambert:2009qw}
  N.~Lambert and P.~Richmond,
  JHEP {\bf 0910}, 084 (2009)
  [arXiv:0908.2896 [hep-th]].




\bibitem{Nastase:2010uy}
  H.~Nastase and C.~Papageorgakis,
  arXiv:1003.5590 [math-ph].
  H.~Nastase and C.~Papageorgakis,
  JHEP {\bf 0912}, 049 (2009)
  [arXiv:0908.3263 [hep-th]].
  H.~Nastase, C.~Papageorgakis and S.~Ramgoolam,
  JHEP {\bf 0905}, 123 (2009)
  [arXiv:0903.3966 [hep-th]].


\bibitem{Cherkis:2008ha}
  S.~Cherkis, V.~Dotsenko and C.~Saemann,
  Phys.\ Rev.\  D {\bf 79}, 086002 (2009)
  [arXiv:0812.3127 [hep-th]].

\bibitem{Gustavsson:2010nc}
  A.~Gustavsson,
  JHEP {\bf 1011}, 043 (2010)
  [arXiv:1008.0902 [hep-th]].


\bibitem{Gomis:2008vc}
  J.~Gomis, D.~Rodriguez-Gomez, M.~Van Raamsdonk and H.~Verlinde,
  JHEP {\bf 0809}, 113 (2008)
  [arXiv:0807.1074 [hep-th]].

\bibitem{Kim:2010mr}
  H.~C.~Kim and S.~Kim,
  arXiv:1001.3153 [hep-th].

\bibitem{Terashima:2008sy}
  S.~Terashima,
  JHEP {\bf 0808} (2008) 080
  [arXiv:0807.0197 [hep-th]].




\bibitem{Madore:1991bw}
  J.~Madore,
  Class.\ Quant.\ Grav.\  {\bf 9}, 69 (1992).

  D.~B.~Fairlie and C.~K.~Zachos,
  Phys.\ Lett.\  B {\bf 224}, 101 (1989).

\bibitem{Lee:2009mm}
  K.~M.~Lee, S.~Lee and S.~Lee,
  JHEP {\bf 0909}, 030 (2009)
  [arXiv:0902.3857 [hep-th]].

\bibitem{Lambert:2010wm}
  N.~Lambert and C.~Papageorgakis,
  JHEP {\bf 1008} (2010) 083
  [arXiv:1007.2982 [hep-th]].

\bibitem{Ho:2008ve}
  P.~M.~Ho, Y.~Imamura, Y.~Matsuo and S.~Shiba,
  JHEP {\bf 0808}, 014 (2008)
  [arXiv:0805.2898 [hep-th]].

\bibitem{Taylor:1996ik}
  W.~Taylor,
  Phys.\ Lett.\  B {\bf 394} (1997) 283
  [arXiv:hep-th/9611042].

\bibitem{Kim:2008gn}
  N.~Kim,
  Phys.\ Rev.\  D {\bf 81}, 086006 (2010)
  [arXiv:0807.1349 [hep-th]].

\bibitem{Antonyan:2008jf}
  E.~Antonyan and A.~A.~Tseytlin,
  Phys.\ Rev.\  D {\bf 79}, 046002 (2009)
  [arXiv:0811.1540 [hep-th]].

\bibitem{Palmkvist}Jakob Palmkvist, private communication.

\bibitem{Mukhi:2008ux}
  S.~Mukhi and C.~Papageorgakis,
  JHEP {\bf 0805}, 085 (2008)
  [arXiv:0803.3218 [hep-th]].

\bibitem{Lambert:2010iw}
  N.~Lambert, C.~Papageorgakis and M.~Schmidt-Sommerfeld,
  arXiv:1012.2882 [hep-th].

\bibitem{Douglas:2010iu}
  M.~R.~Douglas,
  arXiv:1012.2880 [hep-th].

\bibitem{Terashima:2010ji}
  S.~Terashima and F.~Yagi,
  arXiv:1012.3961 [hep-th].

\bibitem{Chen:2010ny}
  C.~H.~Chen, P.~M.~Ho and T.~Takimi,
  JHEP {\bf 1003}, 104 (2010)
  [arXiv:1001.3244 [hep-th]].

\bibitem{Henningson:2004dh}
  M.~Henningson,
  Commun.\ Math.\ Phys.\  {\bf 257} (2005) 291
  [arXiv:hep-th/0405056].

\bibitem{Witten:1996hc}
  E.~Witten,
  J.\ Geom.\ Phys.\  {\bf 22}, 103 (1997)
  [arXiv:hep-th/9610234].

\bibitem{Henningson:1999dm}
  M.~Henningson, B.~E.~W.~Nilsson and P.~Salomonson,
  JHEP {\bf 9909}, 008 (1999)
  [arXiv:hep-th/9908107].


\bibitem{Kim:2008gn}
  N.~Kim,
  Phys.\ Rev.\  D {\bf 81} (2010) 086006
  [arXiv:0807.1349 [hep-th]].

\bibitem{Cherkis:2008ha}
  S.~Cherkis, V.~Dotsenko and C.~Saemann,
  Phys.\ Rev.\  D {\bf 79}, 086002 (2009)
  [arXiv:0812.3127 [hep-th]].

\bibitem{Bashkirov:2010kz}
  D.~Bashkirov and A.~Kapustin,
  arXiv:1007.4861 [hep-th].

\bibitem{Gustavsson:2009pm}
  A.~Gustavsson and S.~J.~Rey,
  arXiv:0906.3568 [hep-th].


\bibitem{Gomis:2008cv}
  J.~Gomis, A.~J.~Salim and F.~Passerini,
  JHEP {\bf 0808}, 002 (2008)
  [arXiv:0804.2186 [hep-th]].

\bibitem{Hosomichi:2008qk}
  K.~Hosomichi, K.~M.~Lee and S.~Lee,
  Phys.\ Rev.\  D {\bf 78}, 066015 (2008)
  [arXiv:0804.2519 [hep-th]].



\end{thebibliography}
\end{document}